\RequirePackage{fix-cm}
\documentclass{article}
\usepackage{arxiv}
\usepackage{lipsum}
\usepackage{placeins}
\usepackage[italicComments=true,indLines=true,noEnd=false]{algpseudocodex}
\usepackage{float}
\usepackage[english]{babel}
\usepackage[utf8]{inputenc} 
\usepackage[T1]{fontenc}    
\usepackage{hyperref}       
\usepackage{url}            
\usepackage{booktabs}       
\usepackage{stackrel,amssymb}
\usepackage{bm}
\usepackage{amsfonts,amsmath,amstext,amsbsy,amsthm}
\usepackage{nicefrac}       
\usepackage[version=4]{mhchem}
\usepackage{microtype}
\usepackage{mathrsfs}
\usepackage{graphicx}
\usepackage[numbers,sort&compress]{natbib}
\usepackage{doi}

\usepackage{xcolor}
\usepackage{enumitem}
\usepackage{caption}
\usepackage[title]{appendix}
\usepackage[capitalise,nameinlink]{cleveref}

\usepackage[bottom]{footmisc}
\usepackage{multirow}
\usepackage{subfigure}
\usepackage{subcaption}
\usepackage{pgfplots}
\usepackage{tikz}

\usepackage{algorithm}
\usepackage{algpseudocodex}
\usetikzlibrary{positioning, arrows.meta}
\newcounter{savesecnumdepth}

\newcommand{\backsection}[2][\backsectionname]{\begingroup\par%
  \small%
  \setcounter{savesecnumdepth}{\value{secnumdepth}}%
  \setcounter{secnumdepth}{0}%
\vskip6pt
\noindent \textbf{#1.} #2\par%
  \setcounter{secnumdepth}{\value{savesecnumdepth}}%
  \endgroup}

\title{Surrogate normal-forms for the numerical bifurcation and stability analysis of Navier-Stokes flows via machine learning}

\author{
\textbf{Alessandro Della Pia\textcolor{blue}{$^{1}$}, Dimitrios G. Patsatzis\textcolor{blue}{$^{1}$}, Gianluigi Rozza\textcolor{blue}{$^{2}$}, Lucia Russo \textcolor{blue}{$^{3,*}$}, Constantinos Siettos\textcolor{blue}{$^{4,}$}\thanks{Corresponding authors, emails: \texttt{lucia.russo@stems.cnr.it, constantinos.siettos@unina.it}}}
{}\\
\textcolor{blue}{$^{(1)}$}Modelling and Engineering Risk and Complexity, \\ \emph{Scuola Superiore Meridionale, School for} \emph{Advanced Studies}, Naples 80138, Italy \\
\textcolor{blue}{$^{(2)}$} SISSA, International School for Advanced Studies, \\
\emph{Mathematics Area, mathLab}, Trieste, 34136, Italy\\
\textcolor{blue}{$^{(3)}$}Institute of Science and Technology for Energy and Sustainable Mobility (STEMS), \\ \emph{Consiglio Nazionale delle Ricerche (CNR)}, Naples 80125, Italy\\
\textcolor{blue}{$^{(4)}$}Dipartimento di Matematica e Applicazioni ‘‘Renato Caccioppoli", \\ \emph{Universit\`a degli Studi di Napoli} \emph{‘‘Federico II"}, Naples 80126, Italy\\
}
\pgfplotsset{compat=1.17}

\begin{document}
\maketitle

\begin{abstract}
Inspired by the Equation-Free paradigm, we propose an ``embed--learn--lift'' framework for constructing minimal-dimensional surrogate ROMs for the numerical analysis of high-fidelity Navier--Stokes simulations, even in the presence of symmetries that standard machine-learning surrogates often fail to preserve.
The framework consists of four main stages. First, manifold learning (here both Proper Orthogonal Decomposition (POD) and Diffusion Maps (DMs)) is used to uncover the intrinsic geometry and dimensionality of the latent space underlying the high-dimensional spatio-temporal Navier--Stokes dynamics across the parameter space. 
In the second stage, we construct ROMs (here, via Gaussian Process regression (GPR)) of minimal dimension---by learning the evolution equations directly on the latent space identified in the first stage. In the third stage, we exploit the toolkit of numerical bifurcation analysis (here MATCONT) to construct bifurcation diagrams and perform systematic stability analysis directly in the latent coordinates. This enables, for example, the efficient continuation of branches of limit cycles emerging from Andronov--Hopf and Neimark--Sacker bifurcations, including the continuation of stable and unstable limit cycles, together with the computation of their periods and stability properties via Floquet multipliers. Such analysis is effectively intractable for the full Navier--Stokes equations.
Finally, by solving the pre-image problem in manifold learning, we reconstruct the bifurcating steady and time-periodic states in the original high-dimensional physical space, thus closing the ``lift'' step of the pipeline. The effectiveness of the proposed methodology is demonstrated on three benchmark two-dimensional configurations: the wake flow past a circular cylinder, the planar sudden-expansion channel flow, and the fluidic pinball. As the Reynolds number $Re$ increases, these systems exhibit Andronov--Hopf, pitchfork (symmetry-breaking), and Neimark--Sacker bifurcations, respectively. We show that DMs-based ROMs allow for a computationally efficient and accurate numerical bifurcation and stability analysis, thus outperforming the widely used POD-ROMs by providing a geometrically consistent parametrization and correctly identifying the intrinsic dimension even in the presence of secondary instabilities, highlighting the need for nonlinear manifold learning methods in CFD.
\end{abstract}


\section{Introduction}
\label{sec:intro}

Numerical simulations of nonlinear phenomena in fluid mechanics have reached today an unprecedented level of detail thanks to the continued advancement of computational algorithms and hardware resources. The complex multiscale spatio-temporal behavior exhibited by fluid flows can be readily reproduced using high-fidelity models, allowing for accurate high-dimensional descriptions of the physical mechanisms at play \cite{haller2025modeling}. 

However, understanding the mechanisms that govern complex phenomena, such as flow instabilities and turbulence, requires more than time-marching simulations: while modern Computational Fluid Dynamics (CFD) codes efficiently integrate transients and even find steady states via Newton solvers, they cannot reliably locate precisely the critical (tipping) points at which instabilities appear or trace the resulting unstable branches. For a systematic and precise analysis of dynamical behaviors, particularly the emergence and evolution of oscillatory patterns (even unstable ones that can explain the mechanisms that lead to turbulence), a systematic bifurcation analysis is indispensable. The numerical bifurcation theory toolkit offers a suite of algorithms and software tools—such as AUTO~\cite{doedel1981auto} and MATCONT~\cite{matcont,govaerts2005numerical}—that facilitate the continuation of both stable and unstable steady states, as well as limit cycles and critical bifurcation points. These packages have been built on decades of dedicated research and development, incorporating extensive expertise in nonlinear dynamics and numerical continuation methods. Their robust algorithms and longstanding validation make them trusted choices for performing accurate bifurcation analysis and continuation, including Hopf, fold, period-doubling and Neimark-Sacker bifurcations, as well as codimension-2 bifurcation tracking for low- to medium-scale systems. However, for high-fidelity CFD simulations, storing and factorizing high-dimensional matrices renders tasks such as limit‐cycle continuation computationally intractable beyond small‐scale test cases~\cite{kavousanakis2008timestepper}. Matrix-free Krylov subspace methods, such as Newton–Krylov solvers, offer a scalable approach for large scale partial differential equation (PDE) systems such as those resulting in fluid flows by avoiding the explicit formation of Jacobians and leveraging efficient matrix-vector products~\cite{kavousanakis2008timestepper,sanchez2004newton,samaey2008newton,Sanchez2002Continuation, sanchez2010multiple,net2015continuation,waugh2013matrix}. In particular, for the Navier-Stokes equations, Newton–Krylov continuation of periodic orbits has been performed in \cite{sanchez2004newton}. However, these methods lack the comprehensive algorithmic capabilities of the aforementioned established bifurcation packages. Such tools provide robust and state-of-the-art techniques for detecting and continuing bifurcations, and offer user-friendly interfaces for systematic analysis. In contrast, matrix-free approaches often require custom implementation and may not support advanced analysis as codimension-2 bifurcation tracking, thus limiting their applicability for in-depth bifurcation studies.

Reduced-order models (ROMs) in  CFD ~\cite{haller2025modeling,temam1995navier,titi1990approximate,quarteroni2007numerical,quarteroni2014reduced,vinuesa_brunton_ML-CFD,hijazi2020data,Pichi} reduce the computational cost and memory demands by orders of magnitude, allowing for faster simulations that can qualitatively capture the emergent behavior, and the systematic bifurcation analysis that with the high-fidelity full-order models would be overwhelming difficult to perform. The fundamental idea behind ROMs is that, despite a complex nature due to the inherent nonlinear behavior, fluid flows often exhibit a few dominant coherent structures, which contain coarse but valuable information about the underlying dynamics over the whole parameter space. A prominent example is the derivation of the Lorenz ordinary differential equations (ODEs) resulting from a low-dimensional truncation of the Navier–Stokes–Boussinesq PDE governing Rayleigh–Bénard convection. To derive such models, one can explicitly enforce knowledge of the high-dimensional governing equations, thus obtaining the so-called intrusive ROMs. A prominent paradigm of ROMs is represented by the so-called normal-forms, namely the simplest possible ROMs of complex/high-dimensional dynamical systems around a bifurcation point \cite{seydel2009practical}. They are constructed by systematically truncating nonessential higher‐order dimensions and terms, thus allowing for a better understanding of the intrinsic dimension of the emergent dynamics, greatly reducing computational cost and facilitating various numerical analysis and control tasks. Normal-forms are of great importance in CFD because they allow complex, high‐dimensional fluid equations—such as the Navier–Stokes system discretized over vast grids—to be rigorously reduced near critical parameter values (e.g., a critical Reynolds number) onto a low‐dimensional model, greatly reducing computational cost. This reduction enables a deeper insight into how and why bifurcations emanate in complex fluid systems, making numerical continuation, stability analyzes, and controller design easier.  

The traditional way to derive such ROMs for fluid flows starts by linearizing the full Navier–Stokes equations around a base state at the bifurcation point and identifying the few critical eigenmodes whose growth rates are near zero. One then uses the invariant/center manifold theory \cite{temam1995navier,hirsch1970invariant,gallay2002invariant,siettos2014equation,wiggins2013normally,siettos2022numerical} or spectral sub-manifold theory \cite{haller2016nonlinear,breunung2018explicit,buza2024spectral} to parameterize a locally invariant, low‐dimensional manifold tangent to those modes—expressed usually as a power‐series, thus truncating higher‐order terms to obtain a self‐consistent reduced ODE system that captures the key nonlinear dynamics around the bifurcation point. The theory of approximate inertial manifolds \cite{temam1995navier,titi1990approximate,Colombo2025ROMHopf} has also been used for the Navier-Stokes equation for the construction of a finite-dimensional invariant manifold onto which all trajectories are exponentially attracted, effectively capturing the long‐term dynamics of the flow. This reduces the infinite‐dimensional PDE to a finite set of ODEs that can approximate the dynamics.

On the other side, data-driven methods have been propelled in recent years by the exponential growth of machine learning (ML) and deep learning methodologies, which aim to construct surrogate ROMs from high-fidelity spatio-temporal simulations~\cite{hijazi2020data,stabile2018finite, Hesthaven_2018,girfoglio2022pod}. The first step involves projecting the high-dimensional spatio-temporal flow dynamics onto a carefully chosen (low-dimensional) latent space. This in CFD is typically done via manifold learning techniques like Proper Orthogonal Decomposition (POD)~\cite{deane1991low,ma2002low,rowley2005model,brunton2020machine}. POD offers a direct, closed-form solution to the pre-image problem, i.e., the reconstruction of the CFD dynamics back to the original high-dimensional space. Nonlinear manifold learning algorithms, such as Diffusion Maps (DMs), have also been exploited ~\cite{coifman2005geometric,coifman2006geometric,nadler2006diffusion,dsilva2018parsimonious,galaris2022numerical,Kevrekidis_DM,Patsatzis_2023,DellaPia_diffusion_2024} to deal with more complex dynamics, including chaotic flows \cite{DellaPia_diffusion_2024}. In these approaches, in contrast to the straightforward closed-form mapping between full and reduced spaces provided by POD, the solution of the pre-image problem suffers from ill-posedness, and different techniques such as geometric harmonics, kernel ridge regression, and the $k$-nearest-neighbors ($k$-NN) algorithm have been proposed~\cite{chiavazzo2014reduced,papaioannou2022time,Patsatzis_2023}. Machine learning-based architectures, such as autoencoders, have also been used to learn a set of coordinates that can parameterize the dynamics of the latent space~\cite{li2020scalable,vlachas2022multiscale,floryan2022data,Eivazi,koronaki2024nonlinear}.
On the part of the construction of surrogate ROMs via machine learning,
various approaches have been proposed including sparse identification of nonlinear dynamical systems (SINDy)~\cite{Brunton_SINDy,Hasegawa_2020,Loiseau_2020,Callaham_2022} coupled also with autoencoders  ~\cite{Champion_SINDy,mentzelopoulos2024variational,CONTI2023116072}, feedforward/recurrent neural networks~\cite{galaris2022numerical,Kevrekidis_DM,bertalan2019learning,lee2020coarse,arbabi2020linking,lee2023learning,dietrich2023learning,fabiani2024task,Srinivasan_2019}, cluster-based networks~\cite{Deng_Noack_2022}, spectral submanifolds~\cite{Cenedese_2022} and Gaussian Process regression models (GPR)~\cite{wan2017reduced,stephenson2018accelerating,ma2021data,ortali2022gaussian}. 
For example,  SINDy has been used to model a range of fluid flows, including laminar and turbulent wakes~\cite{Hasegawa_2020}, convective~\cite{Loiseau_2020} and shear~\cite{Callaham_2022} flows, and it has been also applied in combination with autoencoders to learn latent space coordinates before finding the dynamical system to evolve~\cite{Champion_SINDy,CONTI2023116072}. Deep Neural Network-based Auto Encoders (DNN-AE) have been recently employed to identify reduced-order coordinates of a turbulent flow over a building-like structure on a nonlinear manifold~\cite{Eivazi}. Diffusion Maps and neural networks have been leveraged to build predictive ROMs of a vertically falling liquid film flow in the case of sparse data~\cite{Kevrekidis_DM}. A combination of $\beta$-variational autoencoders and transformers has been used to learn parsimonious and near-orthogonal ROMs for two-dimensional viscous flows in both periodic and chaotic regimes~\cite{rico}.  In \cite{Pichi,HESS2019379,pintore2021efficient},  ROMs and bifurcation diagrams of equilibria are built from steady‐state snapshots of Navier–Stokes simulations, targeting “stable bifurcating branches” using intrusive POD Galerkin projections, with applications to triangular lid-driven cavity and sudden-expansion channel flows. This approach has also been used for the control of solutions on the steady-state unstable branch~\cite{pichi2022driving} and generalized in cases of fluid-structure interaction problems~\cite{khamlich2022model}. In \cite{CONTI2023116072}, POD and Auto Encoders are combined with parametric SINDy to construct low-dimensional models from limited full-order data, enabling efficient bifurcation analysis of dynamical systems, with applications ranging from structural mechanics (clamped–clamped beam) to fluid dynamics (flow past a cylinder).

Here, building on previous efforts rooted in the celebrated Equation-Free (EF) multiscale framework~\cite{kevrekidis2003equation,RUSSO200751} introduced back in 2000's, that is based on the on-demand construction of local models, we propose a four-stage full data-driven framework for the construction of surrogate ``normal-forms'' via machine learning that can approximate not only qualitatively (as the traditional normal-form models) but also quantitatively, with high accuracy, the emergent dynamics of the Navier-Stokes equations, thereby enabling their use in numerical bifurcation and stability analysis, thus dealing also with the presence of continuous symmetries. In particular, in \cite{RUSSO200751}, we showed that a high-fidelity spherical-harmonics simulator of the Smoluchowski PDE for complex  flows of liquid crystals can be exploited to carry out reduced, symmetry-factored bifurcation calculations. The only prerequisites of the EF approach are (i) that the full model is in principle reducible and (ii) a good choice of reduced observables that parametrize an attracting slow manifold; no explicit closure is required because short, appropriately initialized bursts of the full code provide on-demand estimates near that manifold. We also connected this time-stepper approach to the Rowley–Marsden template method to dynamically factor out continuous symmetries (mesoscopic, e.g. liquid-crystal orientation, and macroscopic, e.g. translational/rotational symmetries in flows), making the method applicable to traveling/rotating solutions and to self-similar (scale-invariant) phenomena.

At this point, we use the term ``normal-forms'' to emphasize the central challenge in CFD, namely, determining the correct minimal dimensionality of the latent dynamics across parameter space, highlighting the need of using nonlinear manifold learning (such as DMs) rather than POD, which remains the reduction method most widely used in the field. In addition, such surrogates do not, by default, respect the symmetry properties of the high-fidelity simulator. Consequently, they can break symmetries, producing biased or spurious predictions, thus distorting the bifurcation structure or long-term dynamics. Here, to preserve those symmetries, we use explicit symmetry-aware functions as proposed in \cite{Kevrekidis_PRE}.
Based on the fully data-driven constructed surrogates, one can then exploit the full arsenal of state-of-the-art continuation toolboxes such as MATCONT \cite{matcont,dhooge2006matcont} and AUTO \cite{doedel1981auto}, to perform computationally efficient and highly accurate bifurcation and stability analysis. Such investigations would be computationally intractable in the high-dimensional state space of the full Navier–Stokes simulator, where one typically relies on linear (or weakly nonlinear) stability analysis of equilibria and local theoretical normal forms to trace bifurcation diagrams~\cite{Sierra}.

For our illustration, the proposed methodology is demonstrated on three suitable two-dimensional test-bed  configurations: the wake flow past a circular cylinder, the planar sudden-expansion channel flow, and the fluidic pinball. As the Reynolds number $Re$ increases, these systems undergo qualitatively distinct transitions: the cylinder wake exhibits an Andronov--Hopf bifurcation, the sudden-expansion flow undergoes a pitchfork (symmetry-breaking) bifurcation, and the fluidic pinball displays a Neimark--Sacker bifurcation. Importantly, for the selected flow configurations, the Andronov--Hopf and pitchfork bifurcations constitute primary bifurcations, as they represent the first qualitative transitions encountered by the flow upon variation of $Re$. In contrast, the Neimark--Sacker bifurcation of the fluidic pinball is a secondary bifurcation, arising on a previously established limit-cycle branch generated by a primary instability. In the first step, we apply both POD and DMs to detect the intrinsic dimensionality of the high-dimensional Navier--Stokes numerical simulations. In the second stage, we construct surrogate models in the latent space (here Gaussian Process regression (GPR) models). Then, based on the constructed ROMs, we exploit the toolkit of numerical bifurcation analysis (here MATCONT) to perform numerical bifurcation and stability analysis. Finally, by solving the pre-image problem, we reconstruct the bifurcation diagrams in the original high-dimensional physical space. For all the selected benchmark configurations, the proposed methodology enables the construction of a single surrogate ``normal-form''-like model over the entire range of the bifurcation parameter, allowing for the accurate reconstruction of the full bifurcation diagram, including the continuation of steady states, limit cycles, and invariant tori together with their stability. Moreover, the selection of the test cases allows to demonstrate that nonlinear manifold learning via Diffusion Maps successfully reveals the minimal latent space required to model the flow dynamics in the presence of more complex (e.g.\ secondary) bifurcations, whereas POD suffices only for simpler, primary bifurcation scenarios.

The rest of the paper is organized as follows. In Section~\ref{sec:methodology}, we state the problem and describe the four stage methodology. We begin by briefly presenting the parsimonious DMs algorithm and the GPR modelling approach, which also enables uncertainty quantification (details are given in Appendix~\ref{app:insights}). We then briefly describe the bifurcation and stability analysis techniques that we used focusing on the continuation of steady states and limit cycles observed in the Navier--Stokes simulations. In Section ~\ref{sec:layout}, we describe the three benchmark flow configurations. Results are discussed in Section~\ref{sec:results}, and conclusions are finally provided in Section~\ref{sec:conclusions}.

\section{Methodology}
\label{sec:methodology}

\begin{figure}
	\centering
\begin{tikzpicture}[
node distance=2.7cm and 2.7cm,
box/.style={draw, minimum width=0.0cm, minimum height=0.0cm, minimum width=5.8cm,align=center},
arrow/.style={->, thick, rounded corners, draw=black!80, >=Latex}
]

\node[box] (TL) {
$\dfrac{\partial u_i}{\partial x_i} = 0$\\
$\dfrac{\partial u_i}{\partial t} + u_j \dfrac{\partial u_i}{\partial x_j} = - \dfrac{\partial p}{\partial x_i} + \dfrac{1}{Re} \dfrac{\partial^2 u_i}{\partial x_j \partial x_j}$
};
\node[box, below=3.33cm of TL,xshift=0cm,minimum width=0cm] (BL) {
	$\dfrac{d a_1}{d t}= g_1( a_1, \dots, a_d, Re)$ \\
	$\vdots $\\
	$\dfrac{d a_d}{d t}=g_d(a_1, \dots, a_d, Re)$
};
\node[right=3.04cm of TL, inner sep=0pt,minimum width=0cm] (TR) {
	\begin{tikzpicture}[scale=1.0]
	\node[anchor=north] at (0.0,0.0) {\includegraphics[width=6.0cm,height=2.5cm]{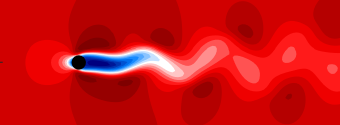}};
	\end{tikzpicture}
};
\node[right=3.92cm of BL, below=2.18cm of TR, inner sep=0pt,minimum width=0cm] (BR) {
	\begin{tikzpicture}[scale=1]
	\node[anchor=north west] at (0,0) {\includegraphics[width=5.5cm,height=4.0cm]{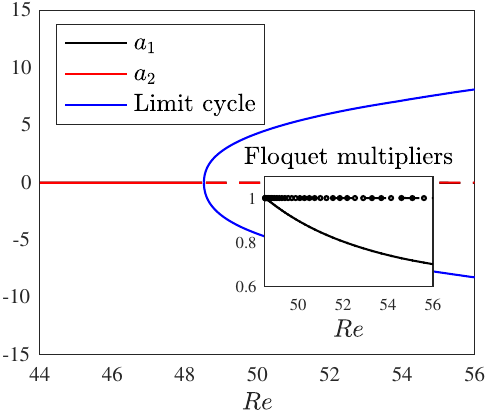}};
	\end{tikzpicture}
};

	\draw[arrow] (TL.south) -- (BL.north) 
node[midway, fill=blue!20, draw=blue!50!black, text=black, font=\small\bfseries, align=center, rounded corners=2pt, inner sep=5pt] {(a) Manifold learning (DMs, POD)\\ (b) Machine learning surrogates (GPR, ANN, SINDy)};
\draw[arrow] (BL.east) -- (BR.west)
node[above=8pt,  xshift=-1.85cm,fill=blue!20, draw=blue!50!black, text=black, font=\small\bfseries, align=center, rounded corners=2pt, inner sep=5pt] {(c) Numerical \\ bifurcation \\ and stability analysis};
\draw[arrow] (BR.north) -- (TR.south)
node[midway,  yshift=-0.1cm, fill=blue!20, draw=blue!50!black, text=black, font=\small\bfseries, rounded corners=2pt, inner sep=5pt] { (d) Pre-image problem ($k$-NN, POD)};

\node[
draw=yellow!50!black,
fill=yellow!20,
rounded corners=2pt,
font=\large\bfseries,
inner sep=6pt
] at ($(TL.north) + (2.6,0.6)$) {High-dimensional physical space};

\node[
draw=yellow!50!black,
fill=yellow!20,
rounded corners=2pt,
font=\large\bfseries,
inner sep=6pt
] at ($(BL.south) - (0.0,0.6)+(2.6,0.0)$) {Low-dimensional latent space};

\end{tikzpicture}
\caption{Four-stage data-driven framework for bifurcation and stability analysis of fluid flows in latent spaces: (a) projection of high-dimensional Navier-Stokes solutions to the latent space (manifold learning, here using Diffusion Maps (DMs) and Proper Orthogonal Decomposition (POD)); (b) modelling of the latent dynamics via machine learning surrogates (here, Gaussian Process regression (GPR)); (c) computation of the latent bifurcation diagram and stability analysis; (d) reconstruction of the physical space solutions (pre-image problem, here using the $k$-NN algorithm and POD).}
\label{fig:overview_framework}
\end{figure}

The pipeline—rooted to the celebrated Equation-Free  (EF) multiscale numerical analysis framework \cite{kevrekidis2003equation}—for the systematic data-driven bifurcation and stability analysis of the Navier-Stokes equations consists of four main steps, shown in Fig.~\ref{fig:overview_framework}: (a) use manifold learning (here both via POD and DMs) to ``restrict/embed'' the high-dimensional spatio-temporal patterns into appropriate latent spaces, (b) learning surrogate ROMs in these latent spaces (here using Gaussian  Process regression, that allows for uncertainty quantification; other surrogates based on neural networks and SINDY \cite{Brunton_SINDy} can be also used), (c) performing numerical bifurcation and stability analysis in the latent spaces, (d) ``lifting'' to the high-dimensional space. A key distinction between our proposed framework and the traditional EF approach that was also coupled with POD for incompressible flows \cite{sirisup2005equation} is that, in the latter, one constructs local mappings and computes the quantities needed for bifurcation analysis via short bursts of microscopic simulations to bridge detailed simulations and emergent dynamics (captured by a few moment variables). In contrast, the proposed method is an extension to the traditional EF framework, in the sense that it learns a model via machine learning from long-term, high-fidelity simulations spanning the entire parameter space. This perspective enables the use of advanced nonlinear manifold-learning techniques, such as Diffusion Maps, and techniques for the solution of the ill-posed pre-image problem, such as geometric harmonics, autoencoders, or the $k$-NN algorithm (here employed) to lift from the latent space to the high-dimensional space. As we have demonstrated in a series of works \cite{Patsatzis_2023, papaioannou2022time, lee2020coarse,fabiani2024task,RUSSO200751, gallos2024data}, this framework can accurately reproduce the emergent bifurcation diagrams for simple one-dimensional PDEs and time series. Here, the framework is exploited to deal with the rich spatio-temporal complexity of the Navier–Stokes PDEs, representing a significant advance, thus extending the EF framework for the bifurcation analysis of high-fidelity simulators introduced in \cite{kevrekidis2003equation,RUSSO200751} including simulators of complex fluids. Here, we show that the proposed pipeline based on DMs, efficiently uncovers a low-dimensional representation that can accurately approximate the rich emergent complex nonlinear dynamics via ROMs, whose governing equations closely resemble normal-forms of the correct dimension. This in turn allows the use of the state-of-the-art numerical bifurcation analysis toolkit, ensuring accurate numerical analysis tasks (such as the continuation of limit cycles and their period, as well as their stability analysis) across a wide range of Reynolds numbers. Actually, the stability analysis and the detection and continuation of solutions emanating from Neimark-Sacker bifurcations is performed, to the best of our knowledge, for the first time, showcasing the potential of the framework to perform advanced numerical bifurcation and stability analysis for the Navier-Stokes PDEs. For the broader CFD community, the framework signals a promising new direction via the use of nonlinear manifold learning such as DMs: the construction of fully data-driven normal-form ROMs of minimal dimensions, offering both high accuracy and computational feasibility. This will facilitate the execution of important tasks in CFD going beyond bifurcation analysis, such as real-time optimization and control.

Here, for our illustrations, we consider the incompressible 2D Navier–Stokes partial differential equations in the dimensionless form: 
\begin{subequations}
\begin{eqnarray}
\dfrac{\partial u}{\partial x} + 	\dfrac{\partial v}{\partial y}&=&0 \label{eq:continuity},
\\
\dfrac{\partial u}{\partial t} + u \dfrac{\partial u}{\partial x} + v \dfrac{\partial u}{\partial y}&=& -\dfrac{\partial p}{\partial x} + \dfrac{1}{Re}\bigg(\dfrac{\partial^2 u}{\partial x^2} + \dfrac{\partial^2 u}{\partial y^2}\bigg) \label{eq:momentum_u},
\\
\dfrac{\partial v}{\partial t} + u \dfrac{\partial v}{\partial x} + v \dfrac{\partial v}{\partial y}&=& -\dfrac{\partial p}{\partial y} + \dfrac{1}{Re}\bigg(\dfrac{\partial^2 v}{\partial x^2} + \dfrac{\partial^2 v}{\partial y^2}\bigg) \label{eq:momentum_v},
\end{eqnarray}
\end{subequations}
where the variables $u\equiv u(x,y,t)$ and $v\equiv v(x,y,t)$ represent the streamwise and normal-to-flow velocity components, respectively, across the 2D spatial domain ($x$, $y$) over time $t$. The variable $p\equiv p(x,y,t)$ denotes the pressure. The governing bifurcation parameter is the Reynolds number $Re$, which represents the relative importance between inertia and viscous effects within the flow. The dimensional variables involved in the definition of $Re$ are case-specific; for the flow configurations considered in this work, $Re$ is given in Eq.~\eqref{eq:Re_cyl} and Eq.~\eqref{eq:Re_coanda} of the following Section~\ref{sec:layout}. 

The representational fidelity of ROMs for accurately describing high-dimensional systems like the Navier-Stokes equations relies on the key assumption that long-term dynamics evolve on a low-dimensional inertial manifold. This manifold is a finite-dimensional, attracting, and invariant subset~\cite{foias1988inertial,constantin1989spectral,constantin2012integral} embedded in the infinite-dimensional phase space of the full-order model. In practice, ROMs are constructed by identifying approximate inertial manifolds (AIMs), which can be parameterized using a small number of dominant modes. A paradigmatic example are the celebrated Lorenz equations, derived as a three-mode reduction of the Rayleigh-Bénard convection equations, which retains the chaotic attractor characteristics of the full-order PDE. Data-driven AIM parameterizations can be constructed numerically using, for example, POD embeddings or numerical analysis-informed nonlinear manifold learning algorithms such as DMs, as recently shown~\cite{DellaPia_diffusion_2024}, but also machine learning-based autoencoders~\cite{koronaki2024nonlinear}, to deal with highly nonlinear phenomena. Such data-driven algorithms complement analytical AIM theory by providing latent coordinates for the slow/long-term dynamics, enabling reduced modeling of complex dissipative systems from data. For a detailed discussion between data-driven ROMs and AIMs, see~\cite{koronaki2024nonlinear}. One of the main aims of this work is to highlight the need for nonlinear manifold-learning methods such as DMs, which, as we show, capture the intrinsic minimal dimensionality required for ROMs and thus enable more demanding numerical-analysis tasks; by contrast, the widely used POD often fails, or requires additional modes that do not reflect the true dimension of the underlying bifurcation structure, thereby missing important dynamical and numerical-analysis insight.

\subsection{The four-stage proposed data-driven framework}
Our aim is to construct a non-intrusive dynamical ROM of minimal dimensions, reflecting a normal-form ROM, by assuming a given set of spatio-temporal snapshots $\{\mathbf{u}_m\}_{m=1,\ldots, M}$ that arise from the discretized Navier-Stokes equations, or from fluid flow experiments. Each snapshot at time $t_m$ contains the velocity variables across the entire spatial domain, which are stacked in the vector $\mathbf{u}_m = [ u(x_i, y_j,t_m), v(x_i,y_j,t_m)]^\top\in \mathbb{R}^{N}$, where $i=1,\ldots,N_x$, $j=1,\ldots,N_y$, and $N=2 N_g$, being $N_g = N_x \times N_y$ the total number of grid points within the spatial domain. Based on the above rationale, we furthermore assume that these points lie on a smooth $d$-dimensional manifold \(\mathcal{M}\subset\mathbb{R}^N\), and we seek a data‐driven mapping:
\begin{equation}
\varPhi:\mathbb{R}^N\;\longrightarrow\;A\subset\mathbb{R}^d,
\quad d\ll N, 
\label{eq:HD2LDmap}
\end{equation}
such that \(\mathbf{a}_m=\varPhi(\mathbf{u}_m)\in \mathbb{R}^d\) captures the dominant coherent structures and emergent dynamics of the flow, where $d$ is an approximation of the intrinsic dimension of $\mathcal{M}$. The construction of this mapping is the first stage of the proposed framework, stage (a) in Fig.~\ref{fig:overview_framework}. 

Toward this aim, one may select for example a linear map \(\varPhi(\mathbf{u}_m)=\mathbf{W}^\top\mathbf{u}_m\), with 
\(\mathbf{W}\in\mathbb{R}^{N\times d}\) containing the leading POD modes, yielding the projection that maximizes the variance in the \(L^2\) sense.  Nonlinear manifold‐learning generalizes this ansatz by constructing \(\varPhi\) to preserve the intrinsic geometry of the data: Isomap, for example, builds a neighborhood graph on \(\{\mathbf{u}_m\}\) and embeds points so that the shortest path distances approximate the manifold geodesics; Diffusion Maps employ Gaussian kernels to define a Markov operator whose leading eigenfunctions embed data such that Euclidean distances in \(\mathbb{R}^d\) approximate diffusion distances in \(\mathcal{M}\); Autoencoder networks learn an encoder \(\varPhi_\theta\) and a decoder \(\Psi_\phi: \mathbb{R}^d \rightarrow \mathbb{R}^N\) by minimizing the reconstruction error \(\sum_{m}\|\mathbf{u}_m - \varPsi_\phi(\varPhi_\theta(\mathbf{u}_m))\|^2\), thus capturing complex nonlinear spatio‐temporal features. Once \(\varPhi\) is obtained, one can subsequently learn sparse surrogate models in the latent space via ML. Hence, whether via linear spectral decompositions or flexible nonlinear embeddings, data‐driven manifold learning provides a rigorous framework for extracting low‐dimensional spatio‐temporal structures of fluid flows, whose full-order models are described by the Navier–Stokes equations.

Having obtained the latent variables 
\(\mathbf{a}(t)=\varPhi(\mathbf{u}(t))\in\mathbb{R}^d\) 
at time \(t\) via the mapping \(\varPhi\), where 
\(\mathbf{u}(t)\in \mathbb{R}^N\) does not necessarily correspond 
to a recorded snapshot \(\mathbf{u}_m\) used to construct \(\varPhi\), 
one can model the governing embedded dynamics in two alternative ways. First, the dynamics can be learned in continuous time as
\begin{equation}
\dfrac{d \mathbf{a}(t)}{dt} \;=\; \boldsymbol{\phi}\bigl(\mathbf{r}(t),\boldsymbol{\chi}\bigr)\;+\;\mathbf{e}(t),
\label{eq:regression_basic}
\end{equation}
or, alternatively, in discrete time as
\begin{equation}
\mathbf{a}(t+T_p) \;=\; \boldsymbol{\phi}\bigl(\mathbf{r}(t),\boldsymbol{\chi}\bigr)\;+\;\mathbf{e}(t),
\label{eq:regression_basic_map}
\end{equation}
where \(d\mathbf{a}(t)/dt\) is the estimated time-derivative of the latent state, 
\(\mathbf{a}(t+T_p)\) is the predicted latent state within the time horizon $T_p$, 
\(\mathbf{r}(t)\in\mathbb{R}^p\) collects the predictors 
(e.g., the current and past latent states or other exogenous inputs), 
\(\boldsymbol{\chi}\in\mathbb{R}^q\) denotes fixed parameters or known forcings, 
and \(\mathbf{e}(t)\) represents unmodeled time-dependent noise. In this work, we have employed both approaches, to ensure compatibility with the continuation toolbox in MATCONT for tracing the various types of bifurcations (see following Section~\ref{sec:results}). Following the first approach (the exact same procedure applies to the second by simply replacing the estimated time-derivative with the predicted latent state), learning a surrogate ROM model in the latent space entails selecting a regression map \(\mathbf{g}:\mathbb{R}^p\times\mathbb{R}^q \times\mathbb{R}^l\to\mathbb{R}^d\) from a hypothesis class \(G\) with parameters \(\boldsymbol{\theta}\in\mathbb{R}^l\), by solving:
\begin{equation}
\min_{\mathbf{g}\in G,\;\boldsymbol{\theta}\in\mathbb{R}^l}\;
\mathcal{L}\!\Bigl(\dfrac{d\mathbf{a}_m}{dt}
,\mathbf{g}\bigl(\mathbf{r}_m,\boldsymbol{\chi};\boldsymbol{\theta}\bigr)\Bigr),
\quad
\dfrac{d\widehat{\mathbf{a}}_m}{dt}
= \mathbf{g}\bigl(\mathbf{r}_m,\boldsymbol{\chi};\boldsymbol{\theta}\bigr),
\label{eq:regOpt}
\end{equation}
over the $M$ observations, where \(\mathcal{L}\) quantifies the prediction error (e.g., mean‐squared error) of the time-derivative estimation. The mapping \(\mathbf{g}\) is the second stage of the proposed framework (stage (b) in Fig.~\ref{fig:overview_framework}), which in practice may be chosen from classes such as SINDy, artificial neural networks (ANN), or GPR models. 

The mapping $\varPhi$ to the latent space, and the ROM governing the dynamics therein, enable making predictions, but more importantly performing system-level tasks in low dimensions, which are infeasible for high-dimensions, such as bifurcation analysis.~Hereby, we perform numerical continuation to track down stable and unstable stationary states of the fluid flow, on the basis of the constructed ROM; stage (c) in Fig.~\ref{fig:overview_framework}. 

Finally, once the surrogate model has produced predictions or stationary states \(\mathbf{a}_*\in\mathbb{R}^d\) in the latent space, we need to solve the pre-image (or “lifting”) problem for reconstructing the corresponding high-dimensional velocity field \(\hat{\mathbf{u}}_*\in\mathbb{R}^N\); stage (d) in Fig.~\ref{fig:overview_framework}.

In general, given the forward embedding by the mapping \(\varPhi:\mathbb{R}^N\to\mathbb{R}^d\), we seek a lifting operator $\varPsi$ that maps new latent points \(\mathbf{a}_*\notin\varPhi(X)\) back to the ambient space.  This solution to the pre‐image (lifting) problem can be formulated as the parameter‐dependent minimization:  
\begin{equation}
\hat{\boldsymbol{c}}
\;=\;
\arg\min_{\boldsymbol{c}}
\Bigl\|\,
\mathbf{a}_*
\;-\;\varPhi\bigl(\varPsi(\mathbf{a}_*;\boldsymbol{c})\bigr)
\Bigr\|_2,
\label{eq:preimage_general}
\end{equation}
subject to any necessary constraints on \(\boldsymbol{c}\), where \(\varPsi(\cdot\,;\boldsymbol{c}):\mathbb{R}^d\to\mathbb{R}^N\) is a lifting operator parametrized by \(\boldsymbol{c}\).  Once \(\hat{\boldsymbol{c}}\) is determined, the high‐dimensional reconstruction is given by:  
\begin{equation}
\hat{\boldsymbol{u}}_*
= \varPsi(\mathbf{a}_*;\,\hat{\boldsymbol{c}}).
\label{eq:LD2HDmap}
\end{equation}
For a review of several approaches for the solution of the inverse problem, see~\cite{chiavazzo2014reduced,papaioannou2022time}.

For the four-stage data-driven framework described above, we have first employed POD to construct the mapping \(\varPhi\) from the high-dimensional state space to the latent space. While POD is a convenient tool for manifold learning since it provides a closed-form solution for the pre-image mapping \(\varPsi\), we found that it fails to identify the correct minimal dimension for more complex bifurcations (see following Section~\ref{sec:results}). To address this, we have further applied parsimonious DMs~\cite{dsilva2018parsimonious,evangelou2022double,DellaPia_diffusion_2024}, coupled with a $k$-NN algorithm \cite{chin2024enabling,Patsatzis_2023} to solve the inherently ill-posed pre-image problem.
Finally, to construct the non-intrusive ROMs in the latent space, we implemented Gaussian Process regression (GPR) models. In the following, we provide details on the data acquisition and discuss each stage of the proposed framework in detail.

\subsection{Numerical solution of the Navier-Stokes equations for data acquisition}
\label{subsec:Navier-Stokes}
For acquiring the high-dimensional data of fluid flow upon which the ROM is constructed, we solve the incompressible Navier-Stokes PDEs in Eqs.~\eqref{eq:continuity}-\eqref{eq:momentum_v}. For each flow configuration, we consider appropriate boundary and initial conditions (see Section \ref{sec:layout} for further details) and perform direct numerical simulations by means of the open-source code BASILISK (\url{http://basilisk.fr}), which implements a second-order accurate finite-volume scheme. As usual for incompressible flows, Eqs.~\eqref{eq:continuity}-\eqref{eq:momentum_v} are solved by means of the so-called projection method, which is here applied on a uniform structured grid of $N_g=N_x \times N_y$ points. In this procedure, a temporary velocity field is first found by ignoring the pressure gradient. In a second step, the temporary field is projected on a space of divergence-free velocity fields by adding the appropriate pressure gradient correction. For a detailed description of the numerical schemes implemented in BASILISK, the reader can refer to~\cite{Popinet2003}. 

To construct the mapping $\varPhi$ in Eq.~\eqref{eq:HD2LDmap} across the values of the bifurcation parameter (i.e., in the pre- and post-bifurcation regimes), we perform $N_{Re}$ direct numerical simulations with varying Reynolds number $Re$ values, uniformly distributed in the range $Re \in [Re_{min}, Re_{max}]$. The resulting numerical solutions consist of spatio-temporal snapshots of the velocity components $u$ and $v$, collected in the column vectors $\mathbf{u}_m\in\mathbb{R}^{2N_g}$. For each numerical solution of a fixed $Re$ value, we record $N_t$ snapshots of $\mathbf{u}_m$ to form the resulting data set, collected in the snapshot matrix $\mathbf{S}=\{\mathbf{u}_m\}_{m=1,\ldots,M}\in \mathbb{R}^{N\times M}$, where $N=2N_g$ and $M=N_t\times N_{Re}$.

\subsection{Reduced-order basis construction}
\label{subsec:basis}

As already discussed, for the first stage of the proposed data-driven framework, we find the mapping $\varPhi$ from the high-dimensional space to the latent space in Eq.~\eqref{eq:HD2LDmap} via manifold learning. For our illustration, we first employ both POD and DMs; for the first two benchmark problems, POD yields a reduced dimension consistent with the dimension of the associated normal-form dynamics. For the third problem, which exhibits more complex nonlinear dynamics, we instead applied DMs as POD with the same dimension fails. In what follows, we briefly summarize both the POD and DM paradigms (see \cite{DellaPia_diffusion_2024} for a more detailed discussion).

\subsubsection{Proper Orthogonal Decomposition}
\label{subsubsec:POD}

For each value of the governing parameter $Re$, we evaluate the fluctuations of the velocity field $\textbf{u}(x,y,t)=[u(x,y,t), v(x,y,t)]^\top$ with respect to the temporal mean $\overline{\textbf{u}}(x,y)$, namely $\textbf{u}'(x,y,t)=\textbf{u}(x,y,t)-\overline{\textbf{u}}(x,y)$. The celebrated POD technique~\cite{Lumley} decomposes these fluctuations as
\begin{equation}
\label{eq:POD_def}
\textbf{u}^\prime(x, y, t)=\sum_{i=1}^{\infty}a_i(t)\boldsymbol{\varphi}_i(x,y),
\end{equation}
where the POD modes $\boldsymbol{\varphi}_i(x,y)$ are mutually orthogonal in space. By defining $\mathbf{S'}=\{\mathbf{u'}_m\}_{m=1,\ldots,M}\in \mathbb{R}^{N\times M}$, the discrete POD modes $\boldsymbol{\varphi}_i$ are obtained via the method of snapshots~\cite{Sirovich}, leading to the following eigenvalue problem for the snapshot covariance matrix $\mathbf{S'}^\top\mathbf{S'}\in\mathbb{R}^{M\times M}$:
\begin{equation} 
\label{eq:eigprobQQt}
\mathbf{S'}^\top\mathbf{S'}\boldsymbol{\psi}_i=\lambda_i \boldsymbol{\psi}_i, \quad i=1, \ldots, M,
\end{equation}
where the POD modes are given by 
\[
\boldsymbol{\varphi}_i  = \frac{\mathbf{S'} \boldsymbol{\psi}_i}{\sqrt{\lambda_i}},
\]
and $\lambda_i\ge 0$ are sorted in descending order, $\lambda_1 \ge \ldots \ge \lambda_M$. By retaining the leading $d \ll M < N$ modes, one obtains a reduced-order POD basis that provides a linear parameterization of the manifold. The high-dimensional fluctuation field is projected onto the corresponding temporal coordinates $a_i(t)$, for $i=1,\dots,d$. Using the POD basis, we define the mapping $\varPhi$ in Eq.~\eqref{eq:HD2LDmap} as 
\[
\mathbf{a}=\varPhi(\mathbf{u})=\mathbf{W}^\top(\mathbf{u}-\overline{\mathbf{u}}),
\]
where $\mathbf{W}=[\boldsymbol{\varphi}_1,\ldots,\boldsymbol{\varphi}_d]\in \mathbb{R}^{N\times d}$ and $\mathbf{a}=[a_1,\ldots,a_d]^\top$ are the leading POD coefficients. The POD projection can be applied to both training snapshots and unseen states in the high-dimensional space, thus providing latent coordinates $\mathbf{a}(t)$ that are subsequently used for the ROM construction via GPR models. We emphasize that the POD basis provides a closed-form expression for the pre-image mapping $\varPsi$ used for reconstruction. In particular, for any latent point $\mathbf{a}_*$, the corresponding high-dimensional field is reconstructed as
\[
\mathbf{\hat{u}}_*=\varPsi(\mathbf{a}_*)=\overline{\mathbf{u}}+\mathbf{W}\mathbf{a}_*.
\]

\subsubsection{Diffusion Maps}
\label{subsubsec:PDMs}

We follow the theoretical formulation and numerical implementation of DMs presented in earlier works \cite{dsilva2018parsimonious,holiday2019manifold,Patsatzis_2023,gallos2024data,DellaPia_diffusion_2024}. 

Assume that the data lie on a smooth, low-dimensional manifold $\mathcal{M} \subset \mathbb{R}^N$. Diffusion Maps then aim to obtain low-dimensional embeddings $\mathbf{a
} \in \mathbb{R}^d$, with $d \ll N$, collected in the matrix $\mathbf{A} \in \mathbb{R}^{M \times d}$, such that Euclidean distances between points $\mathbf{a
}$ approximate the diffusion distances between the original points \cite{nadler2006diffusion}.

The algorithm begins by defining a similarity measure between pairs of data points $\mathbf{u}_i, \mathbf{u}_j \in \mathbb{R}^N$, $\forall i,j=1,\ldots,M$, in the high-dimensional space. Using the Euclidean norm $d_{ij} = \|\mathbf{u}_i - \mathbf{u}_j\|$, we construct a Gaussian kernel $k(\mathbf{u}_i, \mathbf{u}_j)$, which defines the affinity matrix:
\begin{equation}
	\mathbf{F} = [f_{ij}] = [k(\mathbf{u}_i, \mathbf{u}_j)] = \exp\left(-\frac{\|\mathbf{u}_i - \mathbf{u}_j\|^2}{\epsilon^2}\right),
	\label{eq:affinity_matrix}
\end{equation}
where $\epsilon$ controls the local neighborhood size in the high-dimensional space. In our implementation, we set $\epsilon = \mathrm{median}(d_{ij})$, which promotes a relatively large neighborhood. Other strategies for selecting $\epsilon$ exist \cite{singer2009detecting,gallos2021construction}.

Next, the $M \times M$ Markov transition matrix $\mathbf{M}$ is formed by row-normalizing the affinity matrix:
\begin{equation}
	\mathbf{M} = \mathbf{D}^{-1} \mathbf{F}, \quad \text{with} \quad \mathbf{D} = \operatorname{diag}\left(\sum_{j=1}^{M} f_{ij}\right).
	\label{eq:Markovian_matrix}
\end{equation}
Each entry $\mu_{ij}$ of $\mathbf{M}$ represents the probability of moving from point $i$ to point $j$ in the high-dimensional space:
\begin{equation}
	\mu_{ij} = \operatorname{Prob}\left(X_{t+1} = \mathbf{u}_j \mid X_t = \mathbf{u}_i\right).
\end{equation}
Equivalently, using the kernel,
\begin{equation}
	\mu_{ij} = \frac{k(\mathbf{u}_i, \mathbf{u}_j)}{\operatorname{deg}(\mathbf{u}_i)}, \quad \text{with} \quad \operatorname{deg}(\mathbf{u}_i) = \sum_{j=1}^{M} k(\mathbf{u}_i, \mathbf{u}_j),
\end{equation}
recovering Eq.~\eqref{eq:Markovian_matrix}.

The transition matrix $\mathbf{M}$ is similar to the symmetric, positive-definite matrix $\hat{\mathbf{M}} = \mathbf{D}^{-1/2} \mathbf{F} \mathbf{D}^{-1/2}$, which allows an eigendecomposition
\begin{equation}
	\mathbf{M} = \sum_{i=1}^{M} \lambda_i \mathbf{w}_i \boldsymbol{u}_i^\top,
\end{equation}
where $\lambda_i \in \mathbb{R}$ are eigenvalues and $\mathbf{w}_i, \boldsymbol{u}_i \in \mathbb{R}^M$ are left and right eigenvectors, satisfying $\langle \mathbf{w}_i, \mathbf{u}_j \rangle = \delta_i^j$. The right eigenvectors $\boldsymbol{u}_i$ provide an orthonormal basis for the low-dimensional subspace in $\mathbb{R}^d$ spanned by the rows of $\mathbf{M}$, and the best $d$-dimensional approximation is given by the $d$ right eigenvectors corresponding to the $d$ largest eigenvalues.

The standard DMs embedding maps each snapshot $\mathbf{u}_m$ to
\begin{equation}
	\mathbf{a}_m = (\lambda_1 \textit{u}_{1,m}, \ldots, \lambda_d \textit{u}_{d,m}) \equiv (a_{1,m}, \dots, a_{d,m}), \quad m = 1, \ldots, M,
\end{equation}
where $u_{i,m}$ denotes the $m$-th component of the $i$-th right eigenvector corresponding to the $i$-th largest non-trivial eigenvalue $\lambda_i$. This embedding approximates the diffusion distance in the high-dimensional space by Euclidean distance in the embedded space:
\begin{equation}
	D_t^2(\mathbf{u}_i, \mathbf{u}_j) = \left\| \mu_t(\mathbf{u}_i, \cdot) - \mu_t(\mathbf{u}_j, \cdot) \right\|_{L_2, 1/\operatorname{deg}}^2 = \sum_{k=1}^{M} \frac{\left(\mu_t(\mathbf{u}_i, \mathbf{u}_k) - \mu_t(\mathbf{u}_j, \mathbf{u}_k)\right)^2}{\operatorname{deg}(\mathbf{u}_k)},
\end{equation}
with $\mu_t(\mathbf{u}_i, \cdot)$ the $i$-th row of $\mathbf{M}^t$. In our computations, we use $t = 1$.

In practice, the embedded dimension $d$ is determined by the spectral gap of the eigenvalues of the transition matrix  $\mathbf{M}$, assuming that the first $d$ leading eigenvalues are adequate to provide a good approximation of the diffusion distance between all pairs of points \cite{coifman2008diffusion}. However, this is not always the case, since some of the first $d$ eigenvectors may be higher harmonics of previous ones and thus they do not describe new directions along the data set. To consider these cases, we have further employed parsimonious DMs \cite{dsilva2018parsimonious,holiday2019manifold,Patsatzis_2023,gallos2024data} to select the eigenvectors that provide unique directions along the data set, thus providing the best $d$-dimensional embedding. Given the set $\boldsymbol{u}_1,\ldots,\boldsymbol{u}_{i-1}$ of the first $i-1$ DMs eigenvectors, we use a local linear regression model to fit the $i$-th eigenvector $\boldsymbol{u}_i$ against all the previous ones, for each element $m=1,\ldots,M$ as
\begin{equation}
	u_{i,m} \approx \alpha_{i,m} + \boldsymbol{\beta}_{i,m}^\top \mathbf{U}_{i-1,m}, 
\end{equation}
where $\alpha_{i,m}\in\mathbb{R}$, $\boldsymbol{\beta}_{i,m}\in \mathbb{R}^{i-1}$ and $\mathbf{U}_{i-1,m}=[u_{1,m}, \ldots, u_{i-1,m}]^\top$. The parameters $\alpha_{i,m}$ and $\boldsymbol{\beta}_{i,m}$ are found from the solution of the following optimization problem:
\begin{equation}
	(\alpha_{i,m}, \boldsymbol{\beta}_{i,m}) = \underset{\alpha,\boldsymbol{\beta}}{argmin} \sum_{k\neq m} \exp\left(-\dfrac{{\|\mathbf{U}_{i-1,m} - \mathbf{U}_{i-1,k}\|^2}}{\epsilon^2} \right) \left(u_{i,k}-(\alpha + \boldsymbol{\beta}^\top \mathbf{U}_{i-1,k})\right)^2.
\end{equation}
Then, the normalized leave-one-out cross-validation error is measured by the local linear fitting coefficient as
\begin{equation}
	r_i = \sqrt{\dfrac{\sum_{m=1}^M \left(u_{i,m}-(\alpha_{i,m} + \boldsymbol{\beta}_{i,m}^\top \mathbf{U}_{i-1,m})\right)^2}{\sum_{m=1}^M u_{i,m}^2}}.
	\label{eq:LLFc}
\end{equation}
With the above definition, a small or negligible error $r_i$ indicates that the $i$-th eigenvector $\boldsymbol{u}_i$ can be actually predicted from the remaining eigenvectors $\boldsymbol{u}_1, \boldsymbol{u}_2,...,\boldsymbol{u}_{i-1}$ and thus it is a repeated eigendirection (i.e., $\boldsymbol{u}_i$ is considered a harmonic of the previous eigenmodes). Therefore, only the eigenvectors that exhibit a large $r_i$ are selected in a way of seeking the most parsimonious representation.

The resulting DMs embedding is constructed from the retained eigenpairs $\{\lambda_i, \boldsymbol{u}_i\}_{i=1}^d$. The restriction operator (encoder) $\varPhi$, evaluated on a data point $\mathbf{u}_m$, is
\begin{equation}
	\varPhi(\mathbf{u}_m) = (\lambda_1 u_{1,m}, \ldots, \lambda_d u_{d,m}) \equiv (a_{1,m}, \dots, a_{d,m}) = \mathbf{a}_m \in \mathbb{R}^d, \quad m = 1, \ldots, M.
\end{equation}
For new, unseen points, we employ the Nyström method \cite{nystrom1929uber,coifman2006geometric,chiavazzo2014reduced,evangelou2022double,Patsatzis_2023}. The solution of the pre-image problem is achieved via the $k$-nearest-neighbor ($k$-NN) algorithm with convex interpolation~\cite{chin2024enabling}:
\begin{equation}
    \mathbf{\hat{u}}^* = \Psi(\mathbf{a}^*) = \sum_{k=1}^K b_k \mathbf{u}_{S(k)},
\end{equation}
where $\mathbf{u}_{S(k)} \in X$ are the high-dimensional data points whose latent representations $\mathbf{a}_{S(k)}$, known via the DMs embedding, are the $K$ nearest neighbors of the target point $\mathbf{a}^*$ in the embedded space. The convex weights $b_k$ satisfy $\sum_{k=1}^K b_k = 1$ and $b_k \in [0,1]$. In this work, we have used $K=4$.

\subsection{ROMS via Gaussian Process regression}
\label{subsec:GP}

Once the reduced-coordinate embedding of the latent flow dynamics has been obtained, we aim to identify the unknown evolution operator (Eq.~\eqref{eq:regression_basic}) and/or the solution operator (Eq.~\eqref{eq:regression_basic_map}) via GPR. Here, the predictors \(\mathbf{r}(t)\) are taken to be the latent state itself, i.e.\ \(\mathbf{r}(t)=\mathbf{a}(t)\in\mathbb{R}^d\), while the fixed parameters \(\boldsymbol{\chi}\in\mathbb{R}\) include the bifurcation parameter \(Re\). 

The surrogate models can be learned either in continuous time, as ODEs obtained by embedding the differential operator, or in discrete time, as maps obtained by embedding the solution operator; in both cases, the corresponding bifurcation and stability analysis can then be carried out within the reduced space.

Accordingly, we construct a ROM with inputs \(\mathbf{z}(t)=[\mathbf{a}(t),\,Re]^\top \in \mathbb{R}^{d+1}\) and outputs given by the temporal derivatives \(d\mathbf{a}(t)/dt\) in the continuous-time setting, and/or the predicted state $\mathbf{a}(t+T_p)$ within the time horizon $T_p$ in the discrete-time setting. 

The GPR surrogate models for each component of the latent variables $a_i$ for $i=1,\ldots,d$ may be approximated as:
\begin{equation}
\dfrac{da_i}{dt} \;=\; g_i\bigl(\mathbf{a},Re; \boldsymbol{\theta}_i\bigr)\;+\;e_i, \qquad e_i\sim \mathcal{N}(0,\sigma_i^2),
\label{eq:reg_GPs}
\end{equation}
where the time derivative $da_i/dt$ of the $i$-th latent variable is coupled with all other $\mathbf{a}=[a_1,\ldots,a_d]^\top $; the time-dependence is dropped here for conciseness. Each component $g_i(\bm{z};\bm{\theta}_i)=g_i([\mathbf{a}, Re]^\top;\bm{\theta}_i)$ 
in Eq.~\eqref{eq:reg_GPs}, follows an independent (from other components) Gaussian Process $g_i\sim \mathcal{G}\mathcal{P}(0, k_i(\mathbf{z},\mathbf{z}' | \boldsymbol{\theta}_i))$ with zero mean and a kernel $k_i(\cdot)$ parameterized by $\boldsymbol{\theta}_i$ with the inputs $\mathbf{z}$ and $\mathbf{z}'$.

For the implementation of GPR, let us consider the projections $\mathbf{a}_m$  of the $M$ given observations on the latent space, along with the corresponding parameter $Re_m$ values, collected in the matrix $\bm{Z} = [\bm{z}_1, \dots, \bm{z}_M]^\top \in \mathbb{R}^{M \times (d+1)}$, where $\bm{z}_m = [\mathbf{a}_m, Re_m]^\top$. Then, the prior distribution of the $i$-th component $g_i$ in Eq.~\eqref{eq:reg_GPs} is:
\begin{equation}
P(\bm{g}_i \mid \bm{Z}) = \mathcal{N}(\bm{g}_i \mid \bm{0}, \bm{K}_i(\bm{Z}, \bm{Z} \mid \bm{\theta}_i)),
\end{equation}
where $\bm{g}_i = [g_i(\bm{z}_1), \dots, g_i(\bm{z}_M)]^\top \in \mathbb{R}^M$ collects the $i$-th outputs and $\bm{K}_i(\bm{Z}, \bm{Z}) \in \mathbb{R}^{M \times M}$ is the covariance matrix with entries $k_i(\bm{z}_m, \bm{z}_l \mid \bm{\theta}_i)$ for $m,l=1,\ldots,M$.

Let $d\mathbf{a}_i/dt \in \mathbb{R}^M$ denote the vector of time derivatives for the $i$-th latent variable, estimated from the $M$ observed snapshots $\{\mathbf{a}_m\}_{m=1}^M$. Then, predictions at a new point \( \bm{z}_*\in\mathbb{R}^{d+1} \) are made by drawing \( g_i(\bm{z}_*) \) from the joint distribution:
\begin{equation}
\begin{bmatrix}
\frac{d\mathbf{a}_i}{dt} \\
g_i(\bm{z}_*)
\end{bmatrix}
\sim \mathcal{N} \left( \bm{0},
\begin{bmatrix}
\bm{K}_i(\bm{Z}, \bm{Z}\mid \bm{\theta}_i) + (\sigma_i)^2 \bm{I}_M & \bm{k}_i(\bm{Z}, \bm{z}_*\mid \bm{\theta}_i) \\
\bm{k}_i(\bm{z}_*, \bm{Z}\mid \bm{\theta}_i) & k_i(\bm{z}_*, \bm{z}_*\mid \bm{\theta}_i)
\end{bmatrix}
\right).
\end{equation}
It can be shown that the posterior conditional distribution $P(g_i(\bm{z}_*) | d\mathbf{a}_i/dt, \bm{Z}, \bm{z}_*)$ follows a normal distribution $g_i(\bm{z}_*) \sim \mathcal{N}(\mu_{i,*}, \sigma_{i,*}^2)$ with the expected value and variance of the estimation given by:
\begin{align}
\mu_{i,*} &= \bm{k}_i(\bm{z}_*, \bm{Z}\mid \bm{\theta}_i) \left[ \bm{K}_i(\bm{Z}, \bm{Z}\mid \bm{\theta}_i) + \sigma_i^2 \bm{I}_M \right]^{-1} \tfrac{d\mathbf{a}_i}{dt}, \\
\sigma_{i,*}^2 &= k_i(\bm{z}_*, \bm{z}_*\mid \bm{\theta}_i) - \bm{k}_i(\bm{z}_*, \bm{Z}\mid \bm{\theta}_i) \left[ \bm{K}_i(\bm{Z}, \bm{Z}\mid \bm{\theta}_i) + \sigma_i^2 \bm{I}_M \right]^{-1} \bm{k}_i(\bm{Z}, \bm{z}_*\mid \bm{\theta}_i).
\end{align}

The hyperparameters $\bm{\theta}_i$ and noise variance $\sigma_i^2$ in the above equations are estimated by minimizing the negative log marginal likelihood (NLML):
\begin{equation}
-\log P\left(\tfrac{d\mathbf{a}_i}{dt} \mid \bm{Z}, \bm{\theta}_i\right) = \frac{1}{2}\left(\tfrac{d\mathbf{a}_i}{dt}\right)^\top \bm{\Sigma}_i^{-1} \tfrac{d\mathbf{a}_i}{dt} + \frac{1}{2}\log|\bm{\Sigma}_i| + \frac{M}{2}\log 2\pi,
\label{eq:MLE}
\end{equation}
where $\bm{\Sigma}_i = \bm{K}_i(\bm{Z}, \bm{Z}\mid \bm{\theta}_i) + \sigma_i^2 \bm{I}_M$. For our analysis, we have employed a radial basis function kernel with automatic relevance determination (ARD):
\begin{equation}
k_i(\bm{z}_m, \bm{z}_l\mid \bm{\theta}_i) = (\theta_{i,1})^2 \exp\left(-\sum_{j=1}^{d+1} \frac{(z_{m,j} - z_{l,j})^2}{2(\theta_{i,j+1})^2}\right),
\end{equation}
where $z_{m,j}=a_{m,j}$ for $j=1,\ldots,d$ and $z_{m,d+1}=Re_m$, and the hyperparameters $\bm{\theta}_i = [\theta_{i,1}, \dots, \theta_{i,d+2}]$ govern the output scale and the input sensitivities.

Finally, considering the manifold embeddings (from POD or DMs) $\mathbf{a}_m$ and the respective $Re_m$ values, we first estimate the time derivatives $da_{m,i}/dt$ using finite differences. Then, we determine the parameters $\bm{\theta}_i$ of each $g_i$ in Eq.~\eqref{eq:reg_GPs} by minimizing the NLML in Eq.~\eqref{eq:MLE}. The resulting GPR model for the dynamics in the latent space takes the form: 
\begin{equation}
\left\{
\begin{aligned}
\dfrac{d a_1}{d t} &= g_1(a_1, \dots, a_d, Re), \\
&\vdots \\
\dfrac{d a_d}{d t} &=g_d(a_1, \dots, a_d, Re).
\end{aligned}
\right.
\label{eq:DNN_a1}
\end{equation}

In the discrete-time setting instead of estimating time derivatives, we directly consider time-shifted pairs \(\bigl(\mathbf{a}_m(t),\,\mathbf{a}_m(t+T_p)\bigr)\) of the solution operator, and train the regression model to approximate the solution operator in the latent space. In this case, the parameters \(\bm{\theta}_i\) of each map component \(g_i\) in Eq.~\eqref{eq:reg_GPs} are again determined by minimizing the NLML in Eq.~\eqref{eq:MLE}, but using the time-advanced coordinates as targets. The resulting surrogate model for the latent dynamics can be written as the discrete-time map
\begin{equation}
\left\{
\begin{aligned}
a_1(t+T_p) &= g_1(a_1(t), \dots, a_d(t), Re), \\
&\vdots \\
a_d(t+T_p) &= g_d(a_1(t), \dots, a_d(t), Re).
\end{aligned}
\right.
\label{eq:GPR_tilde_a}
\end{equation}

\subsection{Numerical bifurcation analysis}
\label{subsec:BIF}

The identification of the latent-space models in Eqs.~\eqref{eq:DNN_a1}-\eqref{eq:GPR_tilde_a} enables tracking stationary states and assessing their stability, allowing the construction of the coarse-grained bifurcation diagrams. For ROMs learned in the continuous-time setting (Eq.~\eqref{eq:DNN_a1}), the stable stationary states are obtained by numerically finding the roots $\mathbf{a}_*$ of
\begin{equation}
\mathbf{g}(\mathbf{a},Re) = \mathbf{0},
\label{eq:eq_LD}
\end{equation}
using a Newton-Raphson scheme, where $\mathbf{g}=[g_1,\ldots,g_d]$ is the learned latent dynamics operator. For tracking unstable stationary states, one needs to continue branches of varying $Re$ parameter values. This can be achieved with pseudo arc-length continuation, by augmenting Eq.~\eqref{eq:eq_LD} with the condition
\begin{equation}
N(\mathbf{a},Re) = \mathbf{c} \cdot \left(\mathbf{a}-\mathbf{a}_1 \right) + b (Re-Re_1) = 0, \qquad \mathbf{c}\equiv\dfrac{\left(\mathbf{a}_1-\mathbf{a}_0\right)^\top}{\delta s}, \qquad b\equiv \dfrac{Re_1 - Re_0}{\delta s}, 
\label{eq:cont_cond}
\end{equation}
where $(\mathbf{a}_0,Re_0)$ and $(\mathbf{a}_1,Re_1)$ are two previously computed stationary states along the branch and $\delta s$ is the pseudo arc-length step. This condition constrains the next stationary state along the branch to lie on a hyperplane orthogonal to the tangent of the branch at $(\mathbf{a}_1,Re_1)$, with the tangent approximated using the previous point $(\mathbf{a}_0,Re_0)$.

In practice, the next stationary state is obtained by numerically finding the roots of Eqs.~\eqref{eq:eq_LD}-\eqref{eq:cont_cond} via the iterative solution of the linearized $(d+1)$-dimensional system
\begin{equation}
\begin{bmatrix}
\dfrac{\partial \mathbf{g}}{\partial \mathbf{a}} & \dfrac{\partial \mathbf{g}}{\partial Re} \\
\mathbf{c} & b
\end{bmatrix} \begin{bmatrix}
\delta \mathbf{a} \\ \delta Re
\end{bmatrix} = - \begin{bmatrix}
\mathbf{g}(\mathbf{a},Re) \\
N(\mathbf{a},Re)
\end{bmatrix},
\label{eq:cont_sys}
\end{equation}
where $\partial \mathbf{g}/\partial \mathbf{a} \in \mathbb{R}^{d\times d}$ and $\partial \mathbf{g}/\partial Re \in \mathbb{R}^{d}$ are the Jacobian matrix and derivatives of the stationary state condition in Eq.~\eqref{eq:eq_LD}, respectively. At each iteration, Eq.~\eqref{eq:cont_sys} is solved with the Moore-Penrose pseudo-inverse to compute the updates $\delta \mathbf{a}$ and $\delta Re$, until reaching a predefined tolerance criterion. The converged solution $(\mathbf{a}_*,Re_*)$ is then used as the next step along the branch.

Once a stationary state $(\mathbf{a}_*,Re_*)$ is obtained, its stability is determined from the eigenvalues $\sigma_i$ of the Jacobian $\partial \mathbf{g}/\partial \mathbf{a}$: if $\operatorname{Re}(\sigma_i)<0$ for all $i$, the state is stable; if $\operatorname{Re}(\sigma_i)>0$ for any $i$, it is unstable. A bifurcation occurs when $\operatorname{Re}(\sigma_i)=0$ for some $i$. Codimension-1 bifurcations (Hopf, fold, or branching points) mark critical transitions, with Hopf bifurcations giving rise to limit cycles and branching points giving rise to additional equilibrium branches. Codimension-2 bifurcations and detailed criteria can be found in~\cite{kuznetsov1998elements}.
Note that, when dealing with ROMs learned in the discrete-time setting (Eq.~\eqref{eq:GPR_tilde_a}), the analogous stationary states correspond to fixed points of the map:
\begin{equation}
\mathbf{a}_* = \mathbf{g}(\mathbf{a}_*,Re).
\end{equation}
Stability is determined from the eigenvalues of the discrete-time Jacobian: $|\sigma_i|<1$ indicates a stable fixed point.

If a branch of limit cycles emerges, one seeks periodic solutions satisfying $\mathbf{a}(t)= \mathbf{a}(t+T)$, $\forall t$, where $T$ is the period. In this case, $T$ is treated as a parameter and the system in Eq.~\eqref{eq:reg_GPs} is rescaled in time $\tau = t/T$ so that the period is 1. The boundary value problem then reads:
\begin{equation}
\left\{
\begin{aligned}
\dfrac{d\mathbf{a}}{d\tau} - T \mathbf{g}(\mathbf{a},Re) &= 0, \\
\mathbf{a}(0) - \mathbf{a}(1) &= 0,
\end{aligned}
\right.
\label{eq:LC_cont}
\end{equation}
with $T$ unknown, along with the limit cycle state $\mathbf{a}_*$. To remove translational invariance, the system is augmented with a phase condition:
\begin{equation}
\int_0^1 \langle \mathbf{a}(\tau), \dot{\mathbf{a}}_0(\tau) \rangle d\tau = 0, 
\label{eq:phaseC_LC}
\end{equation}
where $\dot{\mathbf{a}}_0$ is the derivative of a previously computed limit cycle. Eqs.~\eqref{eq:LC_cont}-\eqref{eq:phaseC_LC} define a system $\mathbf{G}(\mathbf{a},Re,T)$ analogous to Eq.~\eqref{eq:eq_LD}. Continuation of limit cycles is performed similarly to stationary states, discretizing $\mathbf{a}$ over collocation points. Stability is assessed via Floquet multipliers, which also reveal bifurcation points such as period-doubling, fold, branching, or Neimark-Sacker bifurcations~\cite{kuznetsov1998elements}.

For the implementation of numerical bifurcation analysis of both POD-GPR and DMs-GPR models, we employed MATCONT~\cite{matcont,govaerts2005numerical}, which provides tools for continuation and identification of bifurcation points for stationary states and limit cycles. The learned ROMs from Section~\ref{subsec:GP} (both continuous-time, Eq.~\eqref{eq:DNN_a1}, and discrete-time, Eq.~\eqref{eq:GPR_tilde_a}) are supplied, and derivatives required in Eq.~\eqref{eq:cont_sys} are computed numerically.

MATCONT employs direct numerical solvers for Eq.~\eqref{eq:cont_sys}, which would be intractable for the full Navier-Stokes equations, especially for limit cycles. While matrix-free Krylov methods~\cite{kavousanakis2008timestepper,sanchez2010multiple,net2015continuation,waugh2013matrix} can handle large systems, they lack the full bifurcation analysis capabilities of MATCONT. Our framework performs all continuation and bifurcation analysis in latent space using POD-GPR and DMs-GPR models, and reconstructs solutions in the high-dimensional space via the mapping $\varPsi$ described in Section~\ref{subsec:basis}.

Note that, because of the intrinsic systematic error present in any learned surrogate model, symmetry properties are generally not exactly preserved. Consequently,  the correct identification of e.g., pitchfork bifurcations is restored by enforcing an odd-symmetry transformation of Eq.~\eqref{eq:DNN_a1} prior to performing numerical continuation, as recently proposed by~\cite{Kevrekidis_PRE}. In particular, given the identified right-hand side $g(Re,a_1,a_2,\dots,a_d)$ of the Gaussian Process regression model, we compute
\begin{equation}
\overline{g}(Re,a_1,a_2,\dots,a_d)
=
\dfrac{g(Re,a_1,a_2,\dots,a_d)-g(Re,-a_1,-a_2,\dots,-a_d)}{2}.
\label{eq:GP_transformed}
\end{equation}
The transformed model~\eqref{eq:GP_transformed} is then used to construct the bifurcation diagram, as it will be shown in Section~\ref{subsec:res_hannel}.

Note that symmetry properties are, in general, not known \textit{a priori}. A practical approach is therefore to first compute the bifurcation diagram using the learned surrogate model, and then recompute it after applying the corresponding symmetry transformation. The comparison of the two diagrams can then indicate the presence or absence of the underlying symmetry.

\section{Flow configurations}
\label{sec:layout}

For our illustration, we focus on three different two-dimensional benchmark configurations: the wake flow past a circular cylinder, the sudden-expansion channel flow, and the fluidic pinball (see Fig.~\ref{fig:schematic}). The physical layouts with the associated flow fields and regimes are described in what follows.

\subsection{Wake flow past a circular cylinder}
\label{subsec:cyl}

\begin{figure}
	\centering
	\includegraphics[scale=0.71]{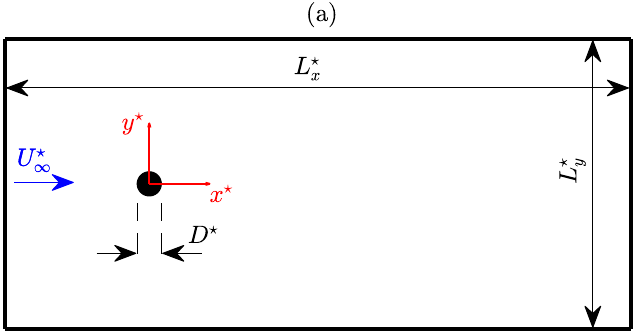}\\
	\vspace{0.25cm} 
    \includegraphics[scale=0.555]{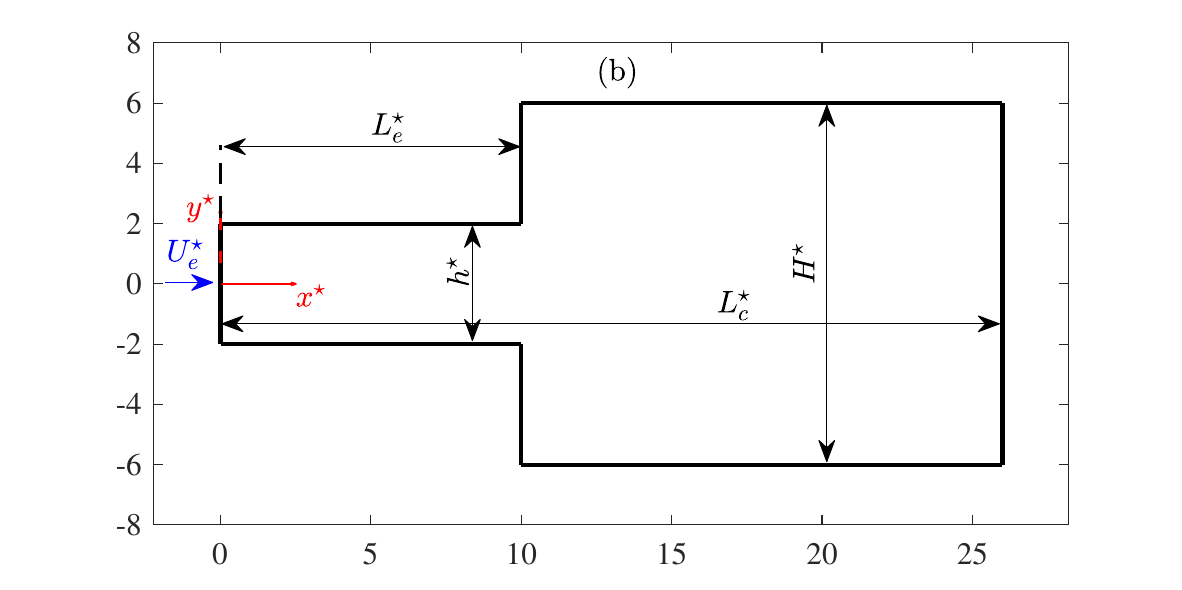}
    \hspace{0.2cm}\includegraphics[scale=0.7]{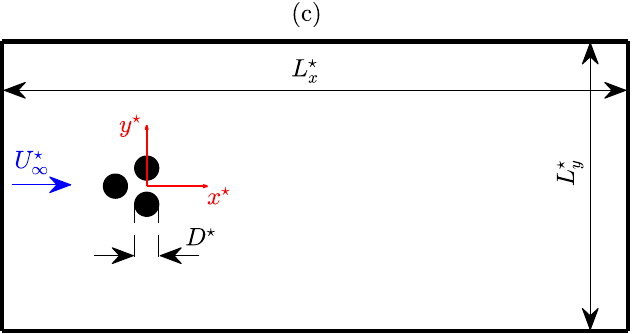}
	\caption{Schematic representations of the three benchmark two-dimensional configurations: wake flow past a circular cylinder (a); planar sudden-expansion channel flow (b); fluidic pinball (c).}
	\label{fig:schematic}
\end{figure}

\begin{figure}[]
	\centering
	\includegraphics[scale=0.8]{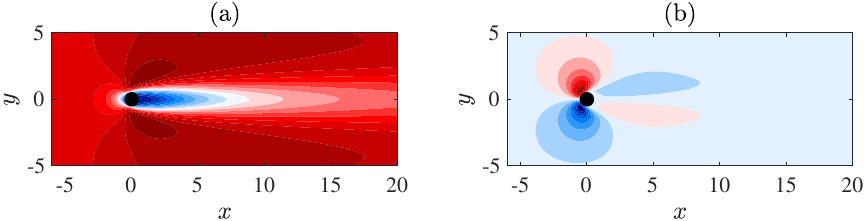}\\
	\vspace{0.25cm}
	\includegraphics[scale=0.8]{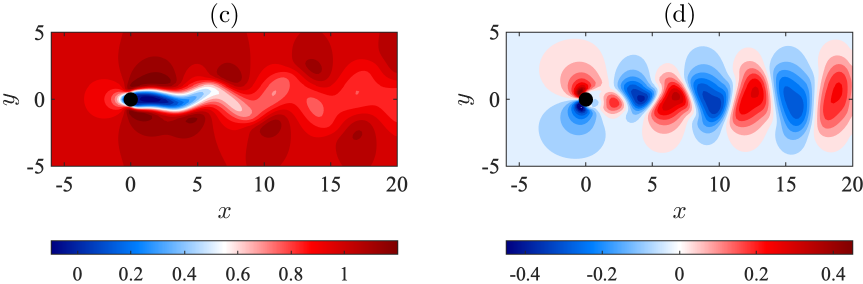}\\
	\vspace{0.25cm}
	\includegraphics[scale=0.8]{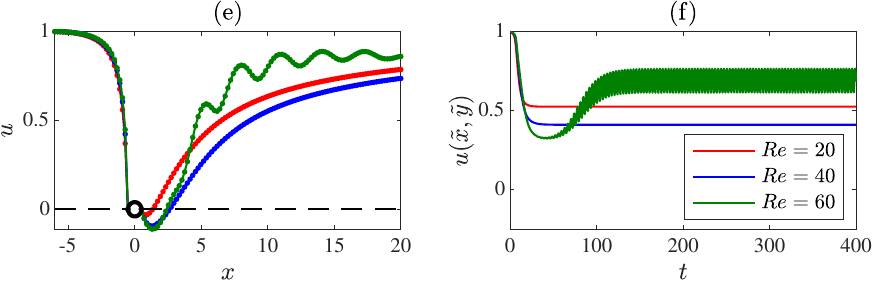}
	\caption{Instantaneous contour maps of $u$ ((a) and (c)) and $v$ ((b) and (d)) velocity components of the cylinder flow for $Re=20$ ((a)-(b)) and $Re=60$ ((c)-(d)). Panel (e) reports the instantaneous spatial distribution of $u$ on the $y=0$ axis for $Re=20$ (red curve), $Re=40$ (blue curve), and $Re=60$ (green curve). Panel (f) shows the temporal evolution of $u$ at the streamwise location $(\tilde{x},\tilde{y}) = (7,0)$ for the same values of Reynolds number. 
	}
	\label{fig:2D_maps}
\end{figure}

The two-dimensional flow around an infinitely long circular cylinder immersed in a uniform stream is schematically represented in Fig.~\ref{fig:schematic}(a). A Cartesian
coordinate system $\mathcal{O}x^\star y^\star$ is placed in the cylinder centre, with the $x^\star$ axis pointing in the flow direction, the $y^\star$-axis along the normal-to-flow direction, and the $z^\star$-axis (not represented) running along the cylinder centreline. The cylinder diameter is denoted as $D^\star$, while the free-stream velocity (aligned with the $x^\star$ direction) is $U^\star_\infty$.
The Reynolds number for this setup is defined as 
\begin{equation}
Re=\dfrac{\rho U^\star_\infty D^\star}{\mu},
\label{eq:Re_cyl}
\end{equation}
where $\rho$ and $\mu$ are the density and dynamic viscosity of the fluid, respectively. Note that all dimensional quantities, except the fluid properties ($\rho, \mu$), are denoted with the superscript $\star$.

Direct numerical simulations of the incompressible Navier–Stokes equations are used to compute the two-dimensional viscous wake behind the cylinder (see previous Section~\ref{subsec:Navier-Stokes} for numerical implementation details). All physical quantities are made dimensionless with respect to the diameter $D^\star$ and velocity $U^\star_\infty$,
\begin{equation}
\label{eq:reference_quantities}
x=\dfrac{x^\star}{D^\star}, \quad  y=\dfrac{y^\star}{D^\star}, \quad u=\dfrac{u^\star}{U^\star_\infty}, \quad v=\dfrac{v^\star}{U^\star_\infty}, \quad p = \dfrac{p^\star}{\rho U^{\star 2}_\infty}, \quad t=t^\star \dfrac{U^\star_\infty}{D^\star}.
\end{equation}
As shown in Fig.~\ref{fig:schematic}(a), the computational domain is a rectangle excluding the interior of the cylinder, with sides equal to $L^\star_x = 26 D^\star$ and $L^\star_y = 12 D^\star$. As the cylinder is considered fixed in the present analysis, a no-slip boundary condition is applied on its surface. At the domain inlet (left side of the rectangle), a uniform velocity profile is prescribed, namely $u=1$ and $v=0$, where $u$ and $v$ are the streamwise and normal-to-flow velocity components, respectively. The same values are also assigned as initial conditions to start the computations.
A standard free-outflow boundary condition is enforced on the domain outlet (right side of the rectangle), while the remaining sides are equipped with homogeneous Neumann boundary conditions for all variables. A uniform structured grid is employed to discretize the physical domain, with mesh size equal to $\Delta x = \Delta y = 0.2$. This corresponds to five grid cells per cylinder diameter, yielding a total number of grid points $N_g= N_x \times N_y = 7800$. Note that such a spatial resolution has been shown to be adequate to reproduce the 2D cylinder wake flow dynamics~\cite{giannetti_2007}.

Spatio-temporal numerical simulations are performed in the range $Re \in [20,60]$. The streamwise $u$ and normal-to-flow $v$ velocity components are stored with a time-step $\Delta t=0.2$, for a total computational time equal to $T=400$. Therefore, 2000 temporal realizations of the velocity field are considered for each value of the Reynolds number.

Snapshots of the simulations are shown in Fig.~\ref{fig:2D_maps} in terms of instantaneous two-dimensional contour maps of the streamwise (panels (a) and (c)) and normal-to-flow (panels (b) and (d)) velocity components by varying the Reynolds number: $Re = 20$ (panels (a)-(b)) and $Re=60$ ((c)-(d)). Moreover, Fig.~\ref{fig:2D_maps}(e) reports the spatial distribution $u(x)$ along the symmetry axis $y=0$ corresponding to $Re=20$, $40$ and $60$, while Fig.~\ref{fig:2D_maps}(f) shows the temporal evolution of $u$ at the streamwise location $(\tilde{x},\tilde{y}) = (7,0)$, for the same $Re$ values.

For relatively low Reynolds number ($Re=20$ and $Re=40$), the flow is steady and characterized by a recirculation wake region, i.e. a flow area where $u<0$. As one can appreciate by comparing red and blue curves in Fig.~\ref{fig:2D_maps}(e), the extent of the wake increases with $Re$. As the Reynolds number increases up to $Re=60$, the velocity field exhibits periodic spatio-temporal oscillations due to the vortex shedding phenomenon originating behind the cylinder (see Fig.~\ref{fig:2D_maps}(c)-(d) and green curves in Fig.~\ref{fig:2D_maps}(e)-(f)).
The onset of the periodic regime is determined by the instabilities developed in the recirculation wake region near $Re_{cr} \approx 49$, which are known to be the manifestation of an Andronov-Hopf bifurcation~\cite{williamson}.

\subsection{Sudden-expansion channel flow}
\label{subsec:channel}
\begin{figure}
	\centering
	\includegraphics[scale=0.85]{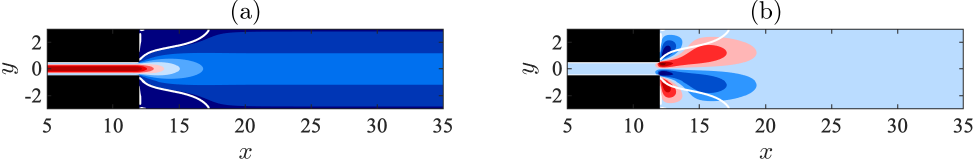}\\
	\vspace{0.25cm}
	\includegraphics[scale=0.85]{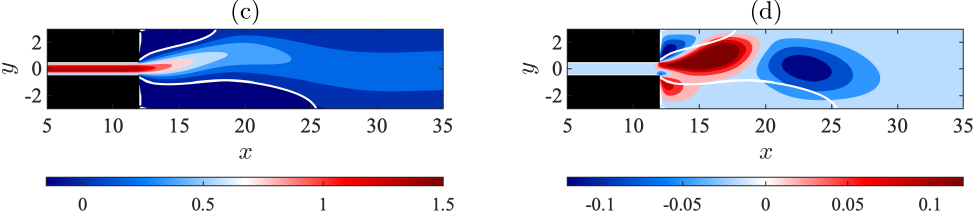}\\
	\vspace{0.25cm}
	\includegraphics[scale=0.8]{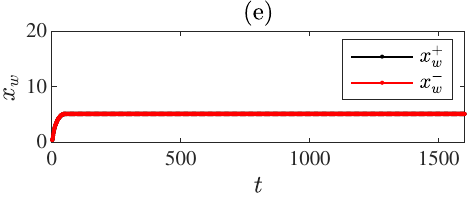}
	\hspace{1.1cm}
	\includegraphics[scale=0.8]{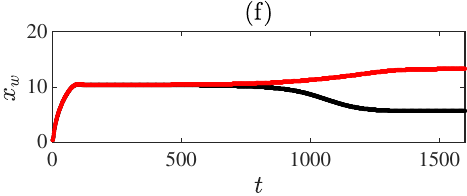}
	\caption{Instantaneous contour maps of $u$ ((a) and (c)) and $v$ ((b) and (d)) velocity components of the sudden-expansion channel for $Re=30$ ((a)-(b)) and $Re=60$ ((c)-(d)). The white curves in panels (a)-(d) denote the flow locations where $u=0$, namely the wake borders. Panels (e)-(f) report the time evolution of the upper ($x^+_w$, black) and lower ($x^-_w$, red) wake region extensions for $Re=30$ and $Re=60$, respectively.}
	\label{fig:2D_maps_BIF}
\end{figure}

The second flow configuration examined is sketched in Fig.~\ref{fig:schematic}(b). It consists of a fluid stream developing in a planar two-dimensional channel, which undergoes a sudden expansion of the cross section as the streamwise direction $x^\star$ increases. The channel has length $L^\star_c$ and the coordinate system $\mathcal{O} x^\star y^\star$ is placed on the symmetry axis at the inlet section. The sudden expansion of the cross-sectional unit area from $A^\star=h^\star$ to $A^\star=H^\star$ is located at $x^\star=L^\star_e$. As in previous works~\cite{Drikakis, quaini}, the expansion ratio $W=H^\star/h^\star$ is set equal to $W=6$, and the channel entrance $L^\star_e$ and total $L^\star_c$ lengths are assigned equal to $L^\star_e=12 h^\star$ and $L^\star_c=60 h^\star$, respectively. These values ensure that the fluid stream development in the entrance region does not affect the flow regimes established in the downstream part of the channel. The Reynolds number for this setup is defined as
\begin{equation}
\label{eq:Re_coanda}
Re=\dfrac{2 \rho U^\star_e h^\star}{\mu},
\end{equation}
where $U^\star_e$ is the averaged velocity at the inlet section, i.e. at $x^\star=0$.

As for the cylinder wake flow configuration, we compute the flow regimes characterizing the sudden-expansion channel flow by direct numerical simulations (implementation details are reported in the previous Section~\ref{subsec:Navier-Stokes}). All physical quantities are made dimensionless with respect to the entrance channel height $h^\star$ and averaged inlet velocity $U^\star_e$,
\begin{equation}
\label{eq:reference_quantities_BIF}
x=\dfrac{x^\star}{h^\star}, \quad  y=\dfrac{y^\star}{h^\star}, \quad u=\dfrac{u^\star}{U^\star_e}, \quad v=\dfrac{v^\star}{U^\star_e}, \quad p = \dfrac{p^\star}{\rho U^{\star 2}_e}, \quad t=t^\star \dfrac{U^\star_e}{h^\star}.
\end{equation}
At the domain inlet (left side of the entrance channel), a parabolic velocity profile with mean value equal to $U^\star_e$ is prescribed. A standard free-outflow boundary condition is enforced on the domain outlet (right side of the channel), while the remaining sides are equipped with no-slip boundary conditions, namely $u=v=0$. The same values are also prescribed as initial conditions to start the computation.
A uniform, structured grid is employed to discretize the physical domain, with mesh size equal to $\Delta x = \Delta y = 0.12$. This corresponds to 9 grid cells within the length scale $h^\star$, namely $N_g= N_x \times N_y = 26214$ total number of grid points, which was necessary to achieve the grid-independence of the flow regimes computed by variation of the Reynolds number $Re$.

Spatio-temporal numerical simulations are performed in the range $Re \in [30,70]$. The streamwise $u$ and normal-to-flow $v$ velocity components are stored with a time-step $\Delta t=0.5$, for a total computational time equal to $T=1600$. Therefore, 3200 temporal realizations of the velocity field are considered for each value of the Reynolds number.

Snapshots of the simulations are shown in Fig.~\ref{fig:2D_maps_BIF} in terms of instantaneous two-dimensional contour maps of the streamwise (panels (a) and (c)) and normal-to-flow (panels (b) and (d)) velocity components by varying the Reynolds number: $Re = 30$ (panels (a)-(b)) and $Re=60$ ((c)-(d)). 

For the lowest Reynolds number considered, a steady symmetric flow with respect to the $y=0$ direction is observed, which is characterized by two recirculation (wake) regions of equal size developing downstream of the expansion (Fig.~\ref{fig:2D_maps_BIF}(a)-(b)). As the Reynolds number increases, the flow symmetry is initially maintained, and the wake regions progressively increase. This aspect is quantified in Fig.~\ref{fig:2D_maps_BIF}(e), which reports the upper $x^+_w$ and lower $x^-_w$ wake regions extension for $Re=30$ as the computational time $t$ increases. Above a critical Reynolds number threshold, denoted here as $Re_{sb}$, the symmetry of the velocity field breaks due to the so-called Coanda effect~\cite{coanda}: a localized increase in velocity near one wall determines a reduction of the local pressure, which establishes a pressure difference maintained across the channel. As a consequence, one of the
downstream recirculation regions expands and the other shrinks, resulting in the steady asymmetric solution observed at $Re=60$ (see Fig.~\ref{fig:2D_maps_BIF}(c)-(d) and Fig.~\ref{fig:2D_maps_BIF}(f)). The symmetry breaking occurs as a
result of a supercritical pitchfork bifurcation in the solution of the Navier–Stokes equations~\cite{battaglia}. In other terms, above $Re = Re_{sb}$ two asymmetric conjugated
stable solutions coexist, while the symmetric solution becomes unstable. Note that the threshold $Re_{sb}$ has been found to be dependent on the channel expansion ratio $W$~\cite{Drikakis,quaini}. For the geometry considered here ($W=6$) and the Reynolds number definition given by Eq.~\eqref{eq:Re_coanda}, the value is $Re_{sb} \approx 44$.

\subsection{Fluidic Pinball}

\begin{figure}
	\centering
	\includegraphics[scale=0.8]{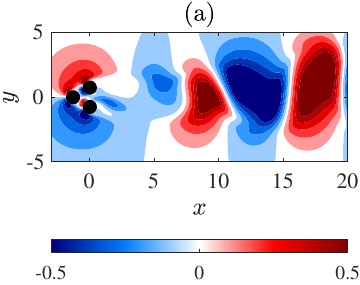}
	\hspace{0.25cm}
	\includegraphics[scale=0.8]{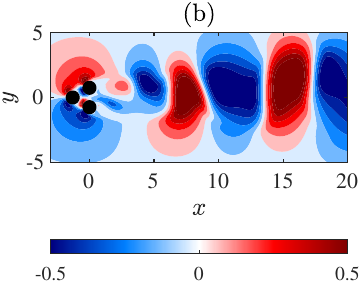}\\
	\vspace{0.5cm}
	\includegraphics[scale=0.8]{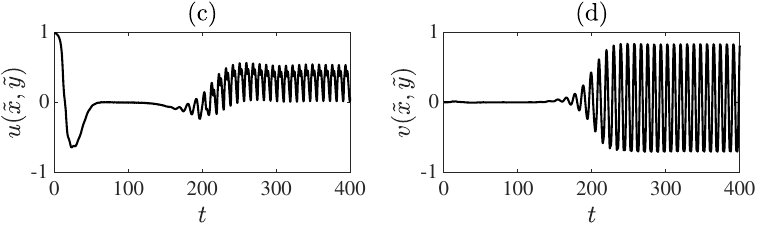}\\
    \vspace{0.5cm}
    \includegraphics[scale=0.8]{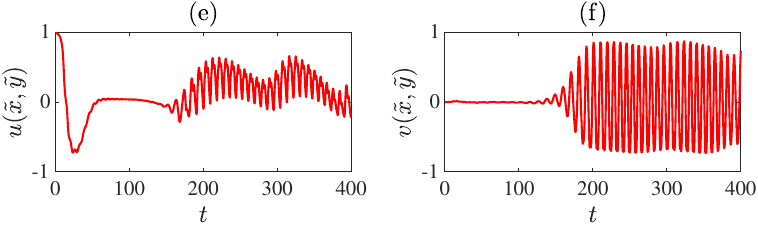}
	\caption{Instantaneous contour maps of $v$ velocity component of the pinball flow for $Re=95$ (a) and $Re=105$ (b).  The temporal evolution of $u$ and $v$ at the streamwise location $(\tilde{x},\tilde{y}) = (7,0)$ is shown for $Re=95$ in panels (c)-(d), and for  $Re=105$ in panels (e)-(f), respectively.
	}
	\label{fig:2D_maps_PIN}
\end{figure}

The last flow configuration examined is sketched in Fig.~\ref{fig:schematic}(c). It consists of three rotatable cylinders of equal diameter $D^\star$, whose axes are located at the vertices of an equilateral triangle. The triangle has a centre-to-centre side length equal to $1.5D^\star$ and is immersed in a viscous, incompressible flow with uniform upstream velocity $U^\star_\infty$. The Reynolds number is defined as in Eq.~\eqref{eq:Re_cyl}. As $Re$ increases, the flow undergoes a sequence of bifurcations—including Andronov--Hopf, pitchfork, and Neimark--Sacker bifurcations—eventually leading to chaotic dynamics \cite{Deng_Noack_2020}. 

As for the cylinder wake flow configuration, we compute the flow regimes characterizing the fluidic pinball by direct numerical simulations (implementation details are reported in the previous Section~\ref{subsec:Navier-Stokes}). All physical quantities are made dimensionless with respect to the cylinders' diameter $D^\star$ and velocity $U^\star_\infty$ (see previous Eq.~\eqref{eq:reference_quantities}). As shown in Fig.~\ref{fig:schematic}(c), the computational domain is a rectangle excluding the interior of the cylinders, with sides equal to $L^\star_x = 26 D^\star$ and $L^\star_y = 12 D^\star$. As the cylinders are considered fixed in the present analysis, a no-slip boundary condition is applied on their surfaces. At the domain inlet (left side of the rectangle), a uniform velocity profile is prescribed, namely $u=1$ and $v=0$, where $u$ and $v$ are the streamwise and normal-to-flow velocity components, respectively. The same values are also assigned as initial conditions to start the computations.
A standard free-outflow boundary condition is enforced on the domain outlet (right side of the rectangle), while the remaining sides are equipped with homogeneous Neumann boundary conditions for all variables. The velocity field is output on the same uniform structured grid used for the circular cylinder flow configuration.

Spatio-temporal numerical simulations are performed in the range $Re \in [90,110]$. The streamwise $u$ and normal-to-flow $v$ velocity components are stored with a time-step $\Delta t=0.2$, for a total computational time equal to $T=400$. Therefore, 2000 temporal realizations of the velocity field are considered for each value of the Reynolds number.

Snapshots of the simulations are shown in Fig.~\ref{fig:2D_maps_PIN} as two-dimensional contour maps of the transverse velocity component, $v$, for two Reynolds numbers: $Re=95$ (panel (a)) and $Re=105$ (panel (b)). Figures~\ref{fig:2D_maps_PIN}(c)–(d) show the temporal evolution of the streamwise and transverse velocity components, $u$ and $v$, at $(\tilde{x},\tilde{y})=(7,0)$ for $Re=95$, while panels (e)–(f) display the corresponding signals for $Re=105$.  
In the Reynolds number range considered, the fluidic pinball undergoes a Neimark--Sacker bifurcation at $Re_{ns} \approx 104$~\cite{Deng_Noack_2020}. This bifurcation destabilizes the periodic spatio-temporal oscillations established in the cylinder wake for $Re < Re_{ns}$, and introduces a low-frequency modulation for $Re > Re_{ns}$, marking the transition to quasi-periodic dynamics. This behavior is associated with slow oscillations of the base-bleeding jet around its deflected mean position behind the cylinders. In agreement with previous literature~\cite{Deng_Noack_2020}, the low-frequency component is roughly one order of magnitude smaller than the primary vortex-shedding frequency, as seen by comparing the pre- and post-bifurcation signals in Fig.~\ref{fig:2D_maps_PIN}(c)–(f).

\section{Results}
\label{sec:results}

The numerical framework presented in Section~\ref{sec:methodology} is here applied to learn the surrogate ROMs of the latent dynamics of the cylinder, channel and pinball flow configurations. Details of the reduced-order basis identification and learning via Gaussian Process regression for the different benchmark cases are reported in Section~\ref{subsec:res_basis} and Appendix~\ref{app:insights}. The identified POD-based and DMs-based ROMs are then employed to perform bifurcation and stability analysis tasks in Sections~\ref{subsec:res_cyl}-\ref{subsec:res_pinball}.

\subsection{Reduced-order basis identification}
\label{subsec:res_basis}

\begin{figure}[]
	\centering
	\hspace{-1.2cm}\includegraphics[scale=0.85]{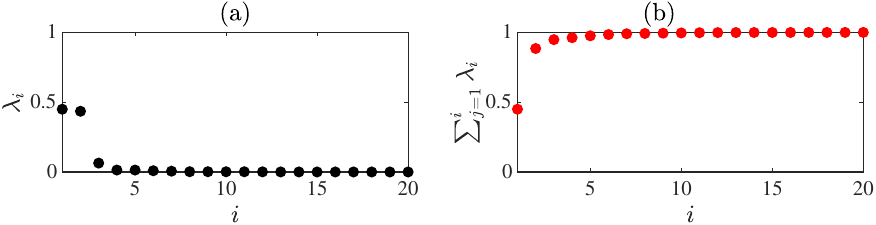}\\
	\vspace{0.25cm}
	\includegraphics[scale=0.85]{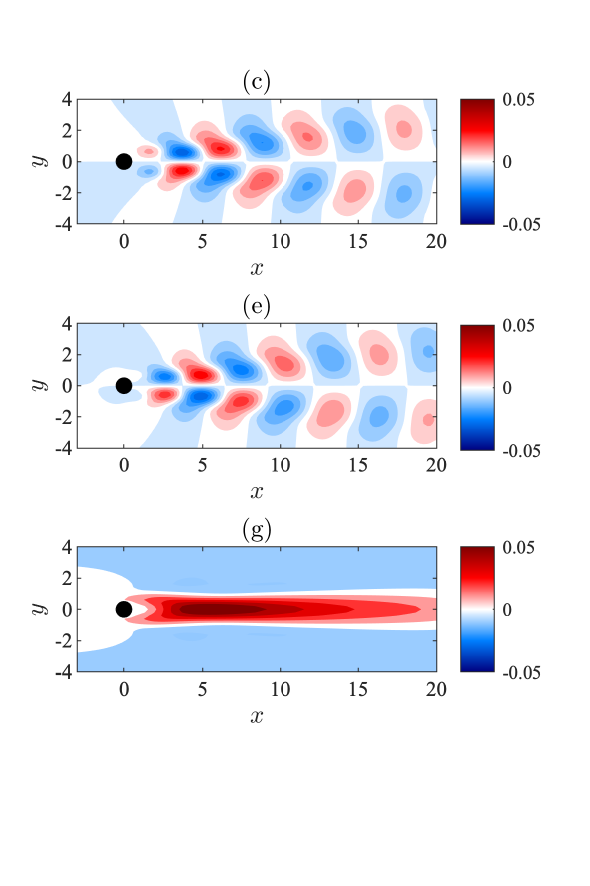}
	\includegraphics[scale=0.85]{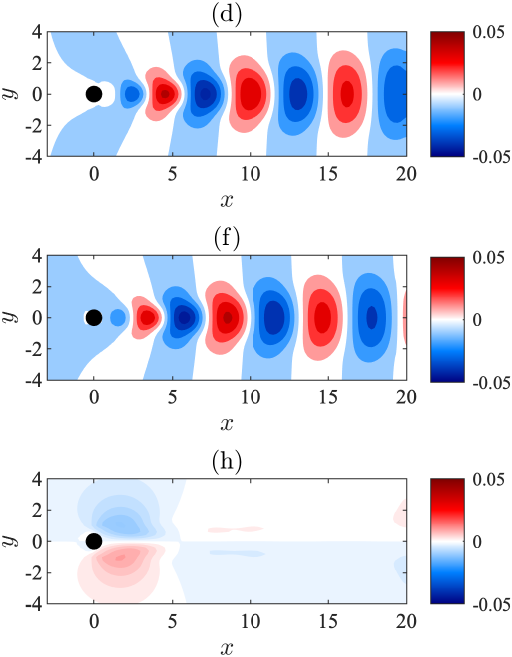}\\
	\vspace{0.25cm}
	\hspace{-0.95cm}\includegraphics[scale=0.85]{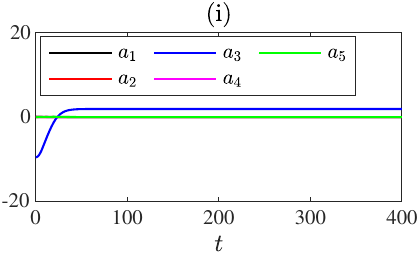}
	\includegraphics[scale=0.85]{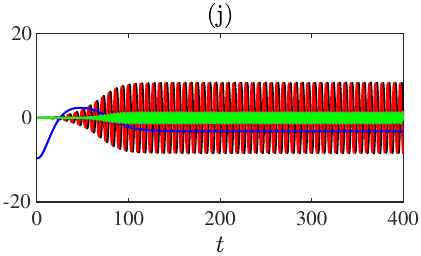}
	\caption{Eigenvalues spectrum and corresponding cumulative sum of the cylinder flow ((a)-(b)). Panels (c)-(d), (e)-(f) and (g)-(h) report the leading spatial modes $\boldsymbol{\varphi}_1$, $\boldsymbol{\varphi}_2$ and $\boldsymbol{\varphi}_3$ for $u$ (left panels) and $v$ (right panels) fields, respectively. Panels (i)-(j) show the temporal evolution of the velocity field projections $a_i$ on the five leading modes $\boldsymbol{\varphi}_i$ ($i=1,\dots,5$) for $Re=40$ and $60$, respectively.}
	\label{fig:POD_modes}
\end{figure}

\begin{figure}[]
	\centering
	\hspace{-1.0cm}\includegraphics[scale=0.85]{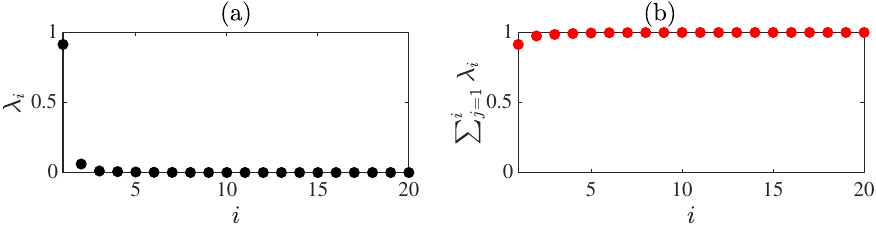}\\
	\vspace{0.25cm}
	\includegraphics[scale=0.85]{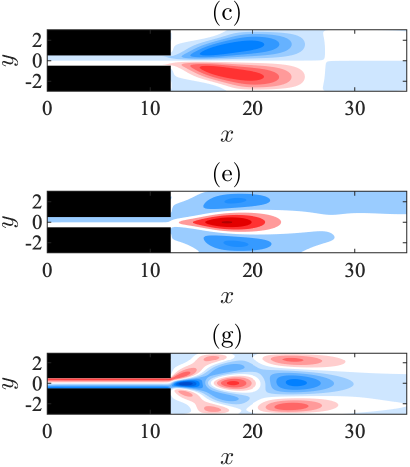}
	\includegraphics[scale=0.85]{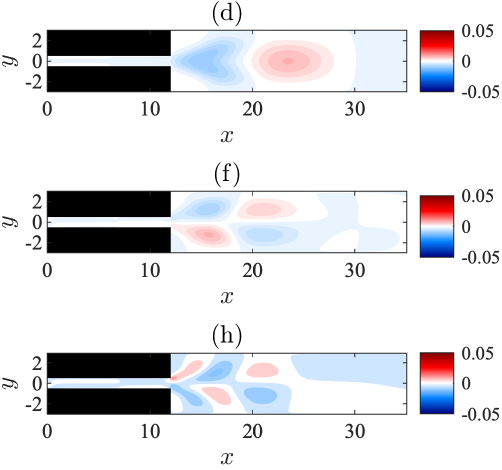}
    \includegraphics[scale=0.8]{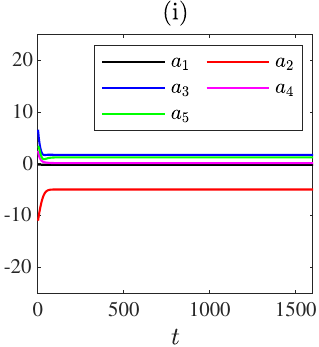}
	\includegraphics[scale=0.8]{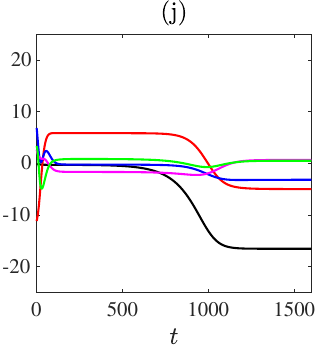}
	\includegraphics[scale=0.8]{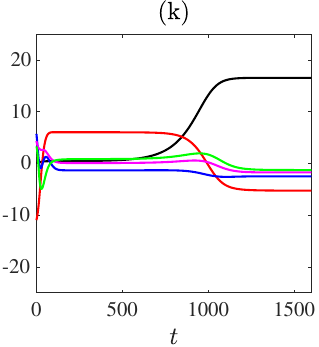}
	\caption{Eigenvalue spectrum and cumulative energy of the sudden-expansion channel flow ((a)–(b)). Leading spatial modes $\boldsymbol{\varphi}_1$, $\boldsymbol{\varphi}_2$, and $\boldsymbol{\varphi}_3$ are shown for the $u$ (left) and $v$ (right) velocity components in panels (c)–(h). Panels (i)–(k) report the temporal evolution of the modal coefficients $a_i$ ($i=1,\dots,5$) at $Re=30$ (i) and $Re=70$, for the upward-deflected (j) and downward-deflected (k) solutions.}
	\label{fig:POD_modes_BIF}
\end{figure}

\begin{figure}[]
	\centering
	\hspace{0.0cm}\includegraphics[scale=0.85]{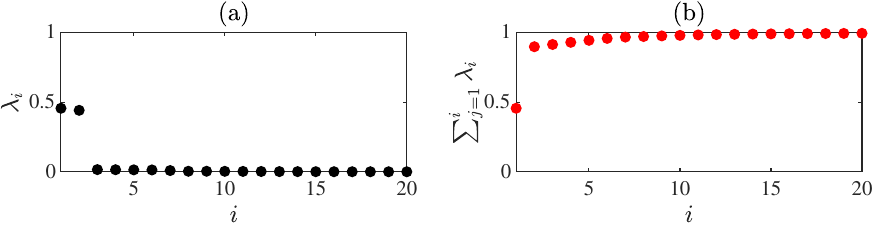}\\
    \vspace{0.25cm}
\hspace{0.0cm}\includegraphics[scale=0.85]{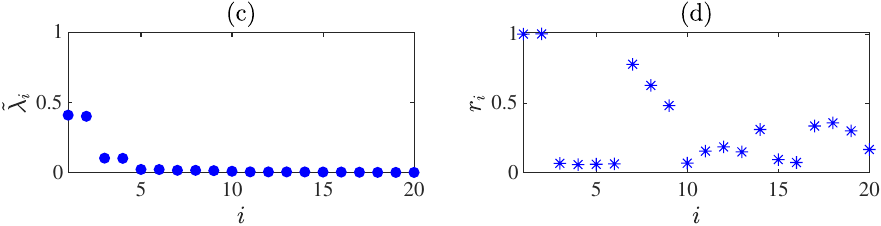}\\
\vspace{0.25cm}
\hspace{0.8cm}\includegraphics[scale=0.85]{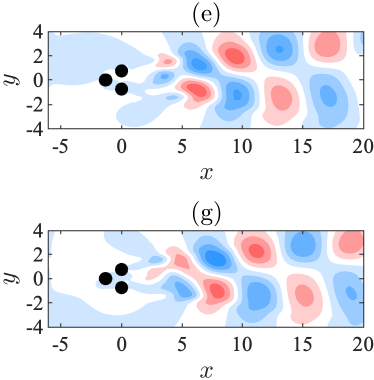}\hspace{0.5cm}
	\includegraphics[scale=0.85]{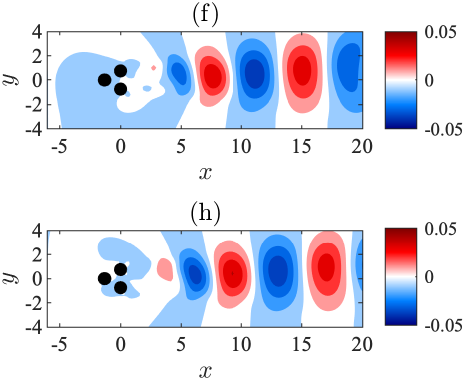}\\
	\vspace{0.25cm}
\hspace{1.2cm}\includegraphics[scale=0.85]{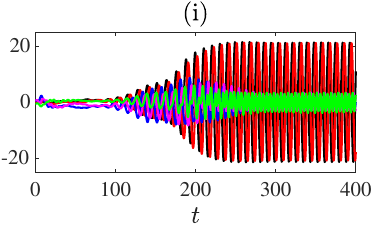}
\includegraphics[scale=0.85]{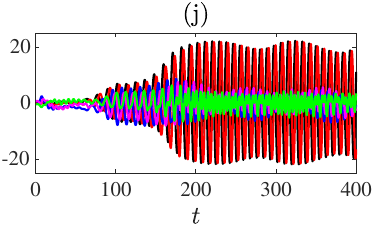}
\includegraphics[scale=0.85]{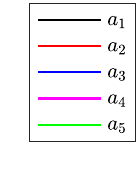}
\caption{\label{fig:POD_and_DMs_spectrum}POD eigenvalues spectrum (a) and corresponding cumulative sum (b) of the pinball flow. Panel (c) reports the DMs eigenvalues (denoted as $\tilde{\lambda_i}$), while panel (d) shows the distribution of the local linear fitting coefficients $r_i$ used for the parsimonious selection. Panels (e)-(f) and (g)-(h) display the leading POD spatial modes $\boldsymbol{\varphi}_1$ and $\boldsymbol{\varphi}_2$ for the $u$ (left) and $v$ (right) velocity fields, respectively. Panels (i)-(j) show the temporal evolution of the velocity field projections $a_i$ on the five leading POD modes $\boldsymbol{\varphi}_i$ ($i=1,\dots,5$) for $Re=95$ and $Re=105$, respectively.}
\end{figure}

The POD eigenvalues spectrum and the corresponding cumulative sum are represented in Fig.~\ref{fig:POD_modes}(a)-(b), Fig.~\ref{fig:POD_modes_BIF}(a)-(b), and Fig.~\ref{fig:POD_and_DMs_spectrum}(a)-(b), respectively for the cylinder, channel and pinball flow configurations. Moreover, the three leading modes $\boldsymbol{\varphi}_1$, $\boldsymbol{\varphi}_2$ and $\boldsymbol{\varphi}_3$ of the cylinder and channel flows are displayed in panels (c)-(d), (e)-(f), and (g)-(h) of Fig.~\ref{fig:POD_modes} and Fig.~\ref{fig:POD_modes_BIF}, respectively, where left panels refer to $u$ and right panels to $v$ velocity fields. The corresponding two leading modes of the pinball configuration are reported in Fig.~\ref{fig:POD_and_DMs_spectrum}, panels (e)-(f) and (g)-(h), respectively.

The projections of the high-dimensional flow field on the five leading modes are shown in Fig.~\ref{fig:POD_modes}(i)-(j), Fig.~\ref{fig:POD_modes_BIF}(i)-(j), and Fig.~\ref{fig:POD_and_DMs_spectrum}(i)-(j) for Reynolds number values both below (left panels) and above (right panels) the critical thresholds $Re_{cr} \approx 49$, $Re_{sb} \approx 44$ and $Re_{ns} \approx 104$.
Values of the parameters employed to build the POD snapshot matrix (see previous Section~\ref{subsec:basis}) corresponding to the cylinder flow are $Re_{min}=40$, $Re_{max}=60$, $N_g=7800$, $N_t=1000$, $N_{Re}=11$, for a total number of snapshots equal to $M=22000$. For the channel flow configuration, we have employed $Re_{min}=30$, $Re_{max}=70$, $N_g=26214$, $N_t=3200$, $N_{Re}=21$, for a total number of snapshots equal to $M=134400$. For the fluidic pinball, we have employed $Re_{min}=90$, $Re_{max}=110$, $N_g=7800$, $N_t=1000$, $N_{Re}=11$, for a total number of snapshots equal to $M=22000$.

As regards the cylinder setup, it can be seen that the first two modes $\boldsymbol{\varphi_1}$ (Fig.~\ref{fig:POD_modes}(c)-(d)) and $\boldsymbol{\varphi_2}$ (Fig.~\ref{fig:POD_modes}(e)-(f)) are physically related to the periodic von K\'arm\'an street of vortices establishing downstream of the cylinder for $Re > Re_{cr}$. The third mode $\boldsymbol{\varphi_3}$ (Fig.~\ref{fig:POD_modes}(g)-(h)) is associated with a shift-mode characteristic of the transient dynamics from the onset of vortex shedding to the periodic von K\'arm\'an wake~\cite{Noack_2003}. The fourth $\boldsymbol{\varphi_4}$ and fifth $\boldsymbol{\varphi_5}$ modes (not shown) represent higher harmonics of the leading two modes, and thus they do not describe new directions along the data set. By looking at the temporal evolution of the velocity field projections $a_i(t)$ for $Re = 40$ (Fig.~\ref{fig:POD_modes}(i)) and $Re=60$ (Fig.~\ref{fig:POD_modes}(j)), it can be clearly appreciated that $\boldsymbol{\varphi_1}$ and $\boldsymbol{\varphi_2}$ represent the minimal number of modes required to describe the bifurcation phenomenon separating the steady and the periodic flow regimes. For this reason, we retain the first $d=2$ modes to train the surrogate Gaussian Process regression model for analyzing the bifurcation in the cylinder flow (see Appendix~\ref{app:insights} for details). Importantly, this demonstrates that the manifold learning method employed—here, POD—automatically identifies the minimal dimensionality needed to capture the bifurcation. As we will see in the more complex pinball test case, such an automatic identification of minimal modes no longer holds.

For the suddenly-expanding channel flow, one can easily observe that the velocity field projection of the first mode $\boldsymbol{\varphi_1}$ (Fig.~\ref{fig:POD_modes_BIF}(c)-(d) and black curves in panels (i)-(k)) embeds the symmetry-breaking bifurcation experienced by the flow as the Reynolds number increases. The temporal evolution of the first latent-dynamics coordinate indeed converges towards a unique fixed-point solution $a_1=0$ for $Re=30$ (Fig.~\ref{fig:POD_modes_BIF}(i)), and to mirrored solutions $a_1=\pm16.5$ for $Re=70$ (Fig.~\ref{fig:POD_modes_BIF}(j)-(k)), namely below and above the threshold $Re_{sb} \approx 44$, respectively. On the other hand, the remaining coordinates $a_i$ ($i = 2, \dots, 5$) exhibit in the long-time limit approximately the same values for the two conjugated solutions existing for $Re > Re_{sb}$ (see red, blue, magenta and green curves in Fig.~\ref{fig:POD_modes_BIF}(j)-(k)). Therefore, only the first $d=1$ mode is retained to learn the surrogate model of the sudden-expansion channel flow dynamics (details are reported in Appendix~\ref{app:insights}), demonstrating again that POD automatically identifies the minimal dimensionality required to reproduce the pitchfork bifurcation.

For the fluidic pinball configuration, the first two modes, $\boldsymbol{\varphi_1}$ (Fig.~\ref{fig:POD_and_DMs_spectrum}(e)-(f)) and $\boldsymbol{\varphi_2}$ (Fig.~\ref{fig:POD_and_DMs_spectrum}(g)-(h)), capture the primary vortex street developing downstream of the three-cylinder arrangement, which is asymmetric due to the upward deflection of the near-wake. For $Re < Re_{ns}$, all modal coefficients exhibit nearly periodic oscillations with a dominant reduced frequency $St = f D/U = \mathcal{O}(10^{-1})$ (Fig.~\ref{fig:POD_and_DMs_spectrum}(i)).

Above the critical Reynolds number $Re_{ns}$, a Neimark–Sacker bifurcation introduces a second, slower timescale, $St = f D/U = \mathcal{O}(10^{-2})$, which modulates the primary vortex-shedding oscillations and produces quasi-periodic dynamics that cannot be captured in a two-dimensional latent space. Hence, selecting $d=2$ modes based solely on the POD spectral gap (Fig.~\ref{fig:POD_and_DMs_spectrum}(a)-(b)) is insufficient to reproduce the observed frequency modulation.
To construct a surrogate, normal-form-like ROM, we therefore apply DMs with parsimonious selection. Analysis of the local linear fitting coefficients $r_i$ (Fig.~\ref{fig:POD_and_DMs_spectrum}(c)-(d)) identifies five dominant coordinates— denoted as $\tilde{a}_1$, $\tilde{a}_2$, $\tilde{a}_7$, $\tilde{a}_8$, and $\tilde{a}_9$—which are retained to train the surrogate Gaussian Process regression model (see Appendix~\ref{app:insights}). This five-dimensional ROM is physically consistent with the Neimark–Sacker bifurcation of the primary limit cycle: the first two coordinates represent the base periodic orbit, an additional pair captures the secondary oscillation generating the invariant torus, and a further coordinate accounts for the neutral direction associated with the time invariance of the orbit. We explicitly note that retaining fewer than five coordinates fails to reproduce the quasi-periodic frequency modulation observed after the bifurcation.

\subsection{Andronov-Hopf bifurcation}
\label{subsec:res_cyl}

\begin{figure}[]
	\centering
	\includegraphics[width=0.9\linewidth]{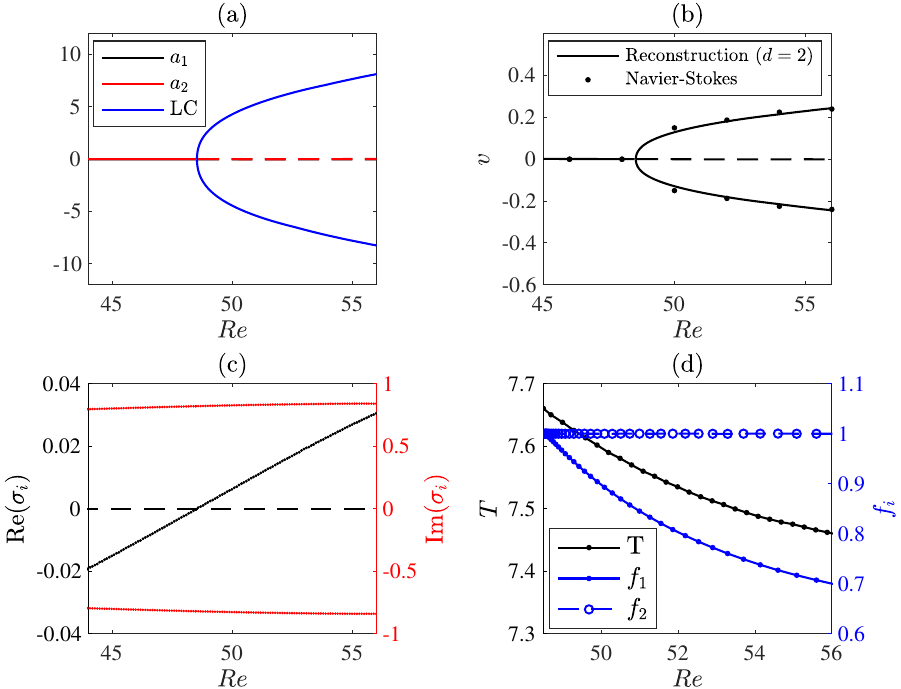}
	\caption{Numerical bifurcation diagram of the POD-based ROM of the cylinder flow by increasing the Reynolds number $Re$ (a). A supercritical Hopf bifurcation occurs at $Re_{cr} = 48.48$: the stable fixed point solution ($a_1 = a_2 = 0$) becomes unstable, giving rise to a branch of limit cycles LC (blue curve). In panel (b), the high-dimensional bifurcation diagram (black continuous curve) is reconstructed in terms of $v$ velocity component at the spatial location $x=10$, $y=0$. Values from the Navier-Stokes solutions are also reported for comparison (black dots).
		Panel (c): a complex-conjugate pair of eigenvalues (red circles) become unstable (i.e., Re($\sigma_i) > 0$ for $i=1,2$) for $Re > Re_{cr}$. Panel (d): limit cycle period $T$ (black curve) and Floquet multipliers $f_1$ and $f_2$ as a function of $Re$ obtained from the limit cycles branch continuation.}
	\label{fig_bifurcation_diagram}
\end{figure}

The numerical integration of the POD-GPR reduced-order model of the cylinder flow dynamics is first performed for two different values of the Reynolds number, spanning both steady laminar ($Re < Re_{cr}$) and periodic vortex shedding ($Re > Re_{cr}$) regimes. Results are shown in Fig.~\ref{fig:DNN_dynamics} in Appendix~\ref{app:insights}, where the time-evolution of the reduced coordinates $a_1$ and $a_2$ in the phase-space is reported for $Re=45$ (panel (a)) and $Re=55$ (panel (b)). Note that values from the test data set, i.e., unseen during the training step, have been employed for the initial conditions $a_1(t=0)$ and $a_2(t=0)$. It can be appreciated that the learned surrogate model qualitatively reproduces the cylinder wake flow dynamics, both in steady and periodic regimes. The temporal evolutions of $a_1$ and $a_2$ reduced coordinates asymptotically converge to the fixed-point stable solution $a_1=a_2=0$ for $Re=45$ (Fig.~\ref{fig:DNN_dynamics}(a)), while periodic limit-cycle oscillations are recovered for $Re=55$ (Fig.~\ref{fig:DNN_dynamics}(b)). 

Once the predictive capability of the POD-based ROM has been validated, a numerical bifurcation diagram is constructed using the continuation method described in Section~\ref{subsec:BIF}. The results are presented in Fig.~\ref{fig_bifurcation_diagram}, where the onset of a Hopf bifurcation is clearly identified at the critical Reynolds number $Re_{cr} = 48.48$.

For $Re < Re_{cr}$, the system exhibits a single stable solution—a fixed-point attractor—represented by the overlapping solid black and red curves $a_1 = a_2 = 0$ in Fig.~\ref{fig_bifurcation_diagram}(a). As the Reynolds number increases beyond $Re_{cr}$, this fixed point becomes unstable to small perturbations, as indicated by the dashed curves in Fig.~\ref{fig_bifurcation_diagram}(a). Therefore, a branch of stable limit cycles (LC) emerges from the Hopf bifurcation point, shown as a blue curve in the same panel. In Fig.~\ref{fig_bifurcation_diagram}(c), the eigenvalues of the Jacobian matrix associated with the system~(\ref{eq:DNN_a1}) are reported as a function of $Re$. As characteristic of a Hopf bifurcation, a complex-conjugate pair of eigenvalues (red curves) cross the imaginary axis and become unstable (i.e., the real part $\mathrm{Re}(\sigma_i) > 0$ for $i=1,2$, black curve) for $Re > Re_{cr}$. Note also that the imaginary part of the eigenvalues, which corresponds to the oscillation frequency of the resulting limit cycles, increases slightly with $Re$ in the range considered here. This trend is consistent with previous findings in the literature~\cite{giannetti_2007}. 

The continuation along the limit cycles branch enables the computation of the oscillation period $T$ (black curve in Fig.~\ref{fig_bifurcation_diagram}(d)) and the associated Floquet multipliers $f_1$ and $f_2$ (blue curves in Fig.~\ref{fig_bifurcation_diagram}(d)), which characterize the stability of periodic solutions. The first Floquet multiplier $f_1$ is equal to one at the bifurcation point and decreases monotonically with increasing $Re$. This behavior reflects the increasing stability of the limit cycle as the Reynolds number grows, since a smaller magnitude of $f_1$ indicates faster decay of perturbations transverse to the periodic orbit. In contrast, the second Floquet multiplier $f_2$ remains constant and equal to 1 throughout the branch. This is an expected trivial result, since $f_2 = 1$ corresponds to the neutral direction associated with the time invariance of the periodic orbit, a characteristic property of limit cycles. The values $f_1 = f_2 = 1$ at the bifurcation point are also consistent with the theoretical prediction for a supercritical Hopf bifurcation.

Finally, by addressing the pre-image problem, the POD-GPR reduced-order solutions can be lifted back to the original high-dimensional physical space to reconstruct the full bifurcation diagram, which is shown in Fig.~\ref{fig_bifurcation_diagram}(b). For comparison, we have also reported the Navier-Stokes solutions (black dots in Fig.~\ref{fig_bifurcation_diagram}(b)) in terms of transverse velocity $v$ stored at the spatial location $x=10$, $y=0$. Note that for $Re > Re_{cr}$, both the maximum and minimum amplitude of $v$ oscillations have been considered at the selected spatial location. It can be seen that the high-dimensional bifurcation diagram is accurately reconstructed by the minimal $d=2$ ROM in the whole Reynolds number range considered, with a maximum relative percentage error equal to 12\% for $Re = 50$.

\subsection{Symmetry-breaking bifurcation}
\label{subsec:res_hannel}

\begin{figure}[]
	\centering
	
	\includegraphics[scale=0.85]{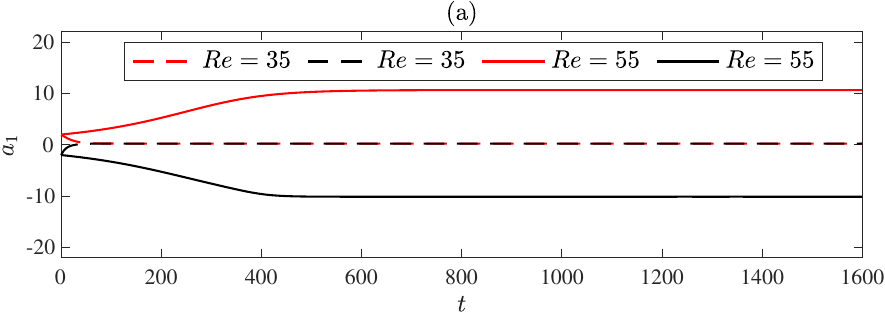}
	\includegraphics[scale=0.85]{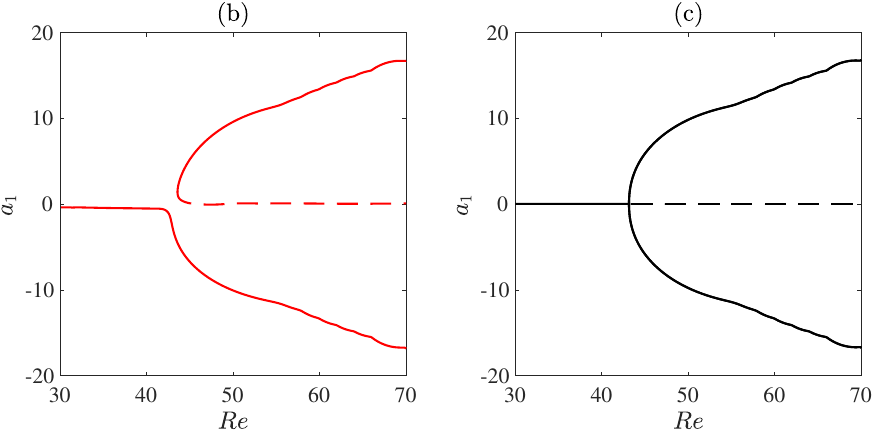}
	\includegraphics[scale=0.85]{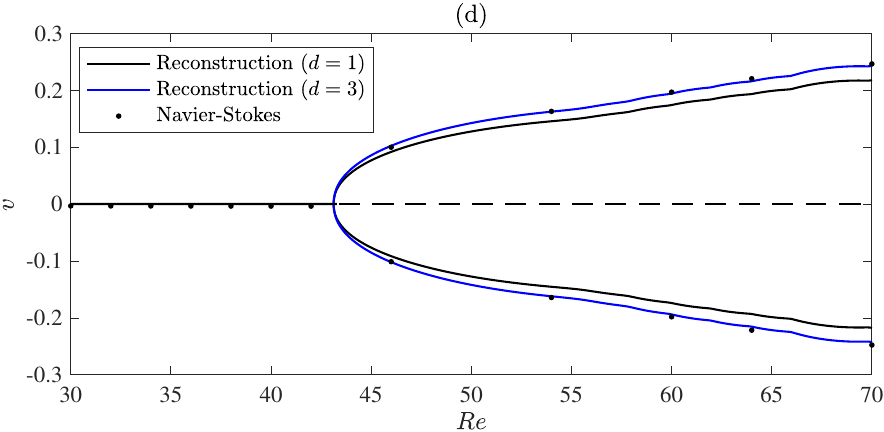}
	\caption{
		Panel (a): reduced-order dynamics of the sudden-expansion channel flow as predicted by time integration of the POD-GPR model by variation of the Reynolds number: $Re=35$ (dashed curves); $Re=55$ (continuous curves). Note that black and red curves refer to different initial conditions $a_1(t=0)$. Panels (b)-(c):
		bifurcation diagram of the sudden-expansion channel flow ROM by increasing the Reynolds number $Re$. In panel (b), the computation is directly performed on the Gaussian Process regression model (Eq.~\eqref{eq:DNN_a1}), while in panel (c) the odd-symmetry transformation (Eq.~\eqref{eq:GP_transformed}) is enforced before computing the solution branches. In panel (d), the high-dimensional bifurcation diagram corresponding to $d=1$ (black continuous curve) and $d=3$ (blue continuous curve) reduced-order coordinates is reconstructed in terms of $v$ velocity component at the spatial location $x=20$, $y=0$. Values from the Navier-Stokes solutions are also reported for comparison (black dots).}
	\label{fig:GP_bifurcation_coanda}
\end{figure}

The numerical integration of the POD-GPR reduced-order model of the sudden-expansion channel flow dynamics is first performed for two different values of the Reynolds number, spanning the pre- ($Re <Re_{sb}$) and post-bifurcation ($Re > Re_{sb}$) regimes. The results are shown in Fig.~\ref{fig:GP_bifurcation_coanda}(a), where the time evolution of the reduced coordinate $a_1$ is reported for $Re=35$ (dashed curves) and $Re=55$ (solid curves). Note that values from the test data set, i.e., unseen during the training step, have been used for the initial condition $a^0_{1} \equiv a_1(t=0)$. Moreover, two conjugated initial conditions have been considered for each $Re$, namely $a^0_{1} = -2$ (black curves) and $a^0_{1} = 2$ (red curves). 

From the analysis of Fig.~\ref{fig:GP_bifurcation_coanda}(a), it can be seen that the learned surrogate model qualitatively reproduces the sudden-expansion channel flow dynamics, both in the pre- and post-bifurcation regimes. Regardless of the assigned initial condition, the temporal evolution of the reduced coordinate $a_1$ converges to the (only) fixed-point stable solution $a_1=0$ for $Re=35$ (dashed curves in Fig.~\ref{fig:GP_bifurcation_coanda}(a)), after a transient time equal to $t \approx 400$. On the other hand, depending on the value of the initial condition $a^0_1$, the upward ($a_1 < 0$) and downward ($a_1 >0$) deflected solutions are asymptotically recovered for $Re=55$ (solid curves in Fig.~\ref{fig:GP_bifurcation_coanda}(a)).

Once the predictive capability of the POD-based ROM has been validated, the numerical bifurcation diagram is first computed by applying to the system~\eqref{eq:DNN_a1} the continuation method described in Section~\ref{subsec:BIF}. As a result, we obtain the bifurcation diagram shown in Fig.~\ref{fig:GP_bifurcation_coanda}(b), which identifies the existence of three steady solutions --- two stable (continuous red curves) and one unstable (dashed red curve) --- for $Re > 43.12$. For $Re < 43.12$, a unique stable steady solution exists. However, the flip symmetry of the generic pitchfork bifurcation is clearly retrieved in the computed diagram. Due to the intrinsic systematic error, which is inherent to any learned surrogate model, the result is a perturbed pitchfork: the lower branch remains permanently stable, while the upper branch exhibits a turning point where a stable and an unstable solution meet. We therefore implement the odd symmetry transformation introduced in Section~\ref{subsec:BIF}, and use the transformed model (Eq.~\eqref{eq:GP_transformed}) to construct the bifurcation diagram shown in Fig.~\ref{fig:GP_bifurcation_coanda}(c), which correctly identifies the symmetry-breaking pitchfork bifurcation at $Re_{sb} = 43.12$.

Finally, by addressing the pre-image problem, the POD-GPR reduced-order solutions can be lifted back to the original high-dimensional physical space to reconstruct the full bifurcation diagram, which is shown in Fig.~\ref{fig:GP_bifurcation_coanda}(d). For comparison, we have also reported the Navier-Stokes solutions (black dots in Fig.~\ref{fig:GP_bifurcation_coanda}(d)) in terms of transverse velocity $v$ stored at the spatial location $x=20$, $y=0$. It can be seen that the high-dimensional bifurcation diagram is already well reconstructed by the minimal $d=1$ ROM (black curve in Fig.~\ref{fig:GP_bifurcation_coanda}(d)) in the whole Reynolds number range considered, with a maximum relative percentage error equal to 22\% for $Re=70$. This is consistent with the expectation that the accuracy of reduced-order surrogates decreases away from the bifurcation point, as normal-form representations are reliable mainly in its vicinity, with deviations growing at larger distances. Nevertheless, the accuracy of the reconstruction improves, reducing the maximum error to 3\% when including the leading $d=3$ reduced-order coordinates in the surrogate POD-GPR model (blue curve in Fig.~\ref{fig:GP_bifurcation_coanda}(d)).

\subsection{Neimark-Sacker bifurcation}
\label{subsec:res_pinball}

\begin{figure}[H]
	\centering
	\includegraphics[scale=.8]{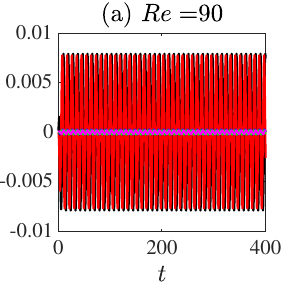}
    \includegraphics[scale=.8]{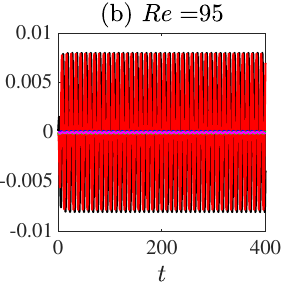}
    \includegraphics[scale=.8]{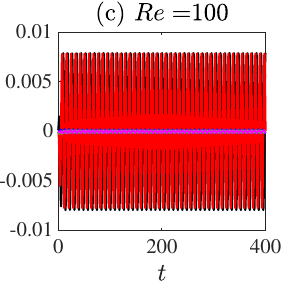}
    \includegraphics[scale=.8]{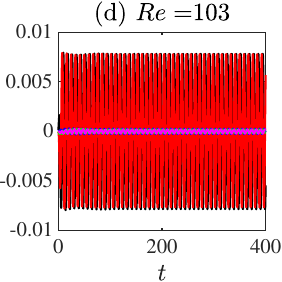}
    \includegraphics[scale=.8]{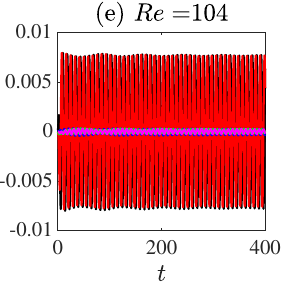}
    \includegraphics[scale=.8]{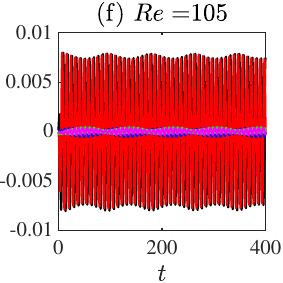}
    \includegraphics[scale=.8]{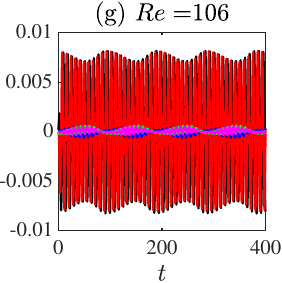}
    \includegraphics[scale=.8]{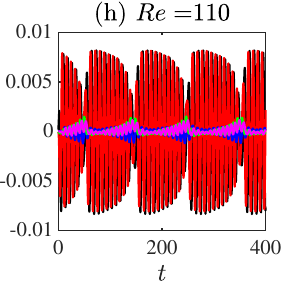}\\
    \vspace{0.20cm}
    \includegraphics[scale=1.0]{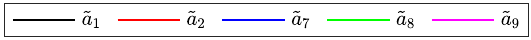}
	\caption{Reduced-order dynamics of the pinball flow as predicted by the DMs-GPR model (Eq.~\eqref{eq:GPR_tilde_a}) by variation of the Reynolds number: $Re=90$ (a); $95$ (b); $100$ (c); $103$ (d); $104$ (e); $105$ (f); $106$ (g); $110$ (h).}
	\label{fig:PIN_dynamics}
\end{figure}

The pinball flow reduced-order dynamics predicted by the DMs-based ROM is shown in Fig.~\ref{fig:PIN_dynamics} for different values of the Reynolds number. As for the previous benchmarks, the initial conditions for the five leading parsimonious DMs coordinates, denoted as $\tilde{a}_i(t=0)$ with $i=1,2,7,8,9$, are taken from the test data set, i.e., from trajectories that were not used during the training stage.

\begin{figure}[H]
	\centering
	\includegraphics[scale=.8]{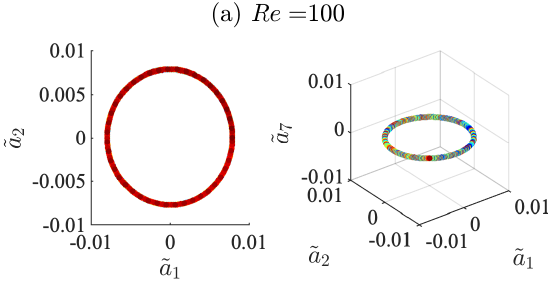}
    \hspace{0.25cm}\includegraphics[scale=.8]{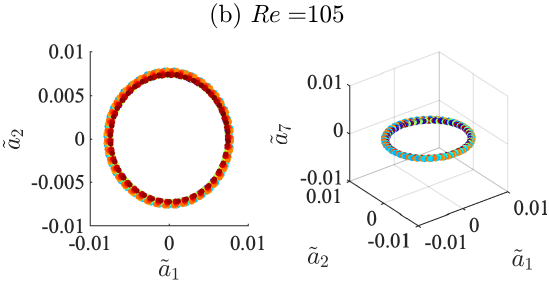}\\
    \vspace{0.5cm}
    \includegraphics[scale=.8]{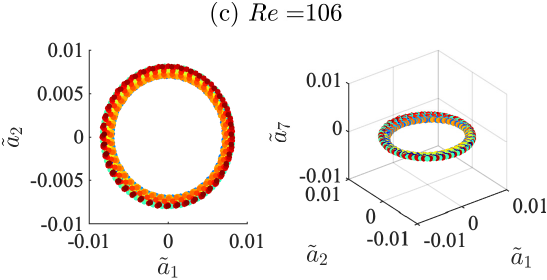}
    \hspace{0.25cm}
    \includegraphics[scale=.8]{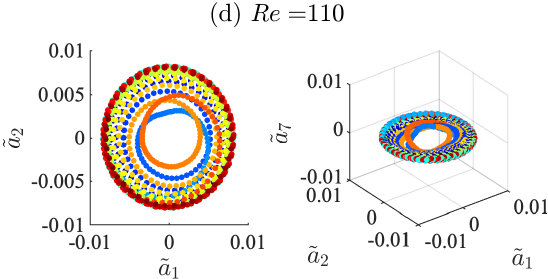}\\
    \vspace{0.5cm}
    \includegraphics[scale=.9]{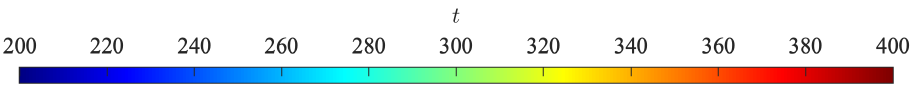}
	\caption{Reduced-order dynamics of the pinball flow in the latent space spanned by the leading parsimonious DMs coordinates ($\tilde{a}_1$, $\tilde{a}_2$, $\tilde{a}_7$), colored by time $t$, by variation of the Reynolds number: $Re=100$ (a); $105$ (b); $106$ (c); $110$ (d).}
	\label{fig:PIN_torus}
\end{figure}

The temporal simulations indicate that purely periodic oscillations are maintained up to $Re=104$ (Fig.~\ref{fig:PIN_dynamics}(a)--(e)). In this regime, the reduced-order dynamics converges to a stable limit cycle characterized by a single dominant frequency associated with the primary vortex-shedding mechanism. For $Re>104$, however, a slow modulation of the main oscillation frequency clearly emerges (Fig.~\ref{fig:PIN_dynamics}(f)--(h)). This modulation reflects the activation of a secondary low-frequency component and marks the transition to quasi-periodic dynamics. In Fig.~\ref{fig:PIN_torus}, the ROM dynamics is further illustrated for $Re=100$ (panel (a)), $105$ (b), $106$ (c), and $110$ (d), both in the 2D ($\tilde{a}_1$,$\tilde{a}_2$) and in the 3D ($\tilde{a}_1$, $\tilde{a}_2$, $\tilde{a}_7$) embedding space defined by the leading parsimonious DMs coordinates. For $Re=100$, the attractor is a limit cycle, as expected for periodic vortex shedding. As the Reynolds number increases and the secondary frequency develops, the attractor undergoes a topological transition: the limit cycle progressively deforms into an invariant torus, clearly visible at $Re=106$ (Fig.~\ref{fig:PIN_torus}(c)). At $Re=110$, the trajectories begin to densely explore a larger region of phase space (Fig.~\ref{fig:PIN_torus}(d)), suggesting the close onset of chaotic dynamics. Physically, this transition reflects the increasing interaction among multiple instability mechanisms in the wake of the pinball, leading to amplitude and phase modulations of the primary shedding mode.

\begin{figure}[H]
	\centering
	\includegraphics[scale=.8]{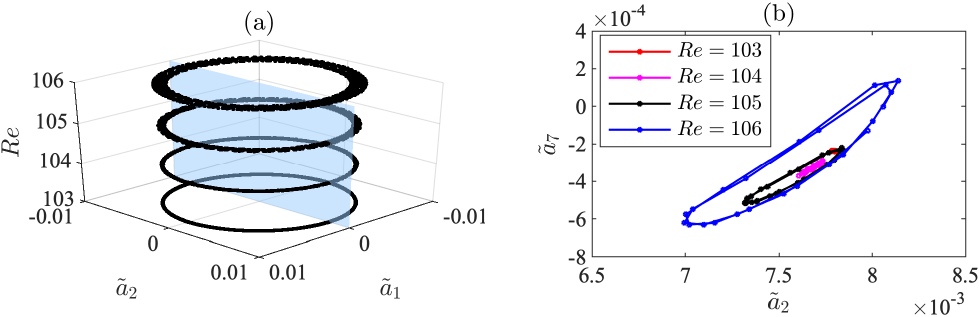}\\
    \vspace{0.1cm}
    \includegraphics[scale=.9]{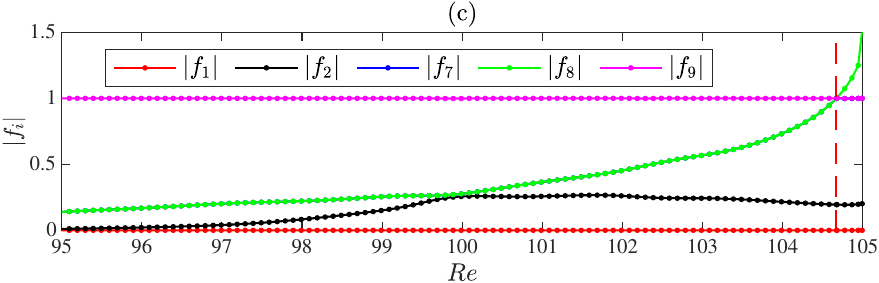}
	\caption{Attracting manifold of the pinball flow dynamics represented by the first two DMs coordinates $(\tilde{a}_1,\tilde{a}_2)$ for different Reynolds numbers $Re$ near the Neimark--Sacker bifurcation point (a), and corresponding Poincar\'e sections in the $\tilde{a}_2$--$\tilde{a}_7$ plane (b). The plane $\tilde{a}_1 = 0$, used to define the Poincar\'e sections, is highlighted in light blue in panel~(a). Panel~(c) reports the Floquet multipliers $f_i$ as a function of the Reynolds number $Re$, as computed via limit-cycle continuation of the DMs--GPR reduced-order model in MATCONT. A Neimark--Sacker bifurcation is detected at $Re_{ns}=104.67$, as indicated by the vertical dashed line (see also Table~\ref{tab:floquet_ns}).}
	\label{fig:PIN_poincare}
\end{figure}

The emergence of the invariant torus is further quantified in Fig.~\ref{fig:PIN_poincare}. Panel (a) shows the attracting manifold of the dynamics projected onto the first two DMs coordinates $(\tilde{a}_1,\tilde{a}_2)$ for different values of the Reynolds numbers close to the Neimark--Sacker bifurcation point. The plane $\tilde{a}_1=0$, used to define the Poincar\'e sections, is highlighted in light blue. The corresponding Poincar\'e sections in the $\tilde{a}_2$--$\tilde{a}_7$ plane are reported in panel (b). The transition from a single fixed point (associated with a stable limit cycle) to a closed invariant curve in the Poincar\'e section provides clear evidence of the birth of a torus in the full three-dimensional embedding space. In other terms, the rise of an invariant torus in the latent space spanned by the leading parsimonious DMs coordinates ($\tilde{a}_1$, $\tilde{a}_2$, $\tilde{a}_7$) corresponds to the appearance of a closed orbit in the Poincar\'e map, as expected for a Neimark--Sacker bifurcation.

\begin{table}[]
\centering
\renewcommand{\arraystretch}{1.3}
\begin{tabular}{c c c c c c }
\hline
$f_1$ & $f_2$ & $f_7$ & $f_8$ & $f_9$ & $T$\\
\hline
$0.0001$ & $0.1949$ & $0.7064 + 0.7082i$ & $0.7064 - 0.7082i$ & $1.0000$ & $8.38$\\
\hline
\end{tabular}
\caption{Floquet multipliers and limit-cycle period of the DMs-GPR reduced-order model at the Neimark--Sacker bifurcation point ($Re_{ns}=104.67$) detected via limit cycle continuation in MATCONT (see also Fig.~\ref{fig:PIN_poincare}).}
\label{tab:floquet_ns}
\end{table}

To accurately identify the bifurcation point, we performed a bifurcation analysis of the DMs--GPR reduced-order model in MATCONT, employing the limit-cycle continuation method described in Section~\ref{subsec:BIF}. The continuation is initialized from the periodic limit-cycle solution obtained at $Re=95$. 

\begin{figure}[H]
	\centering
    \includegraphics[scale=.78]{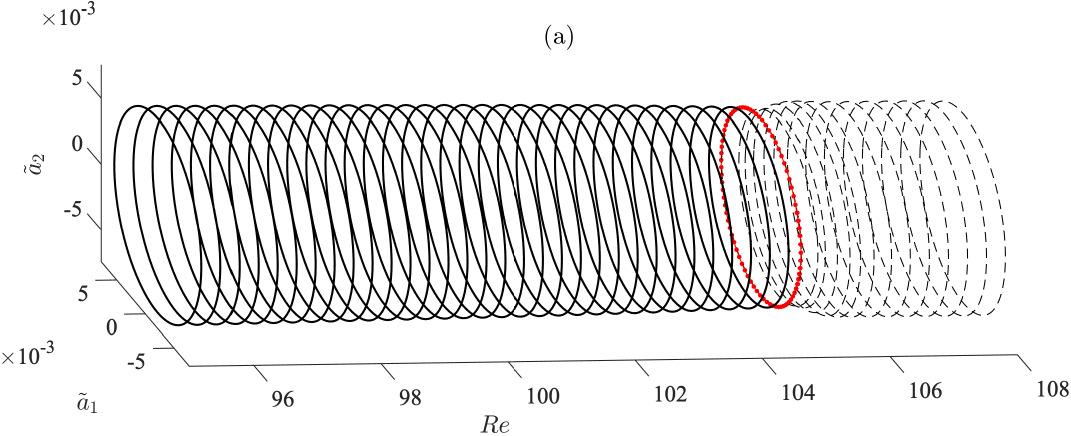}\\
     \vspace{0.5cm}
    \includegraphics[scale=.8]{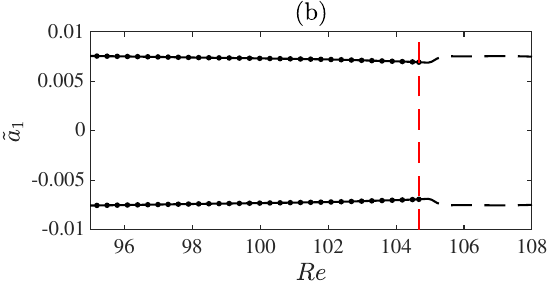}
    \includegraphics[scale=.8]{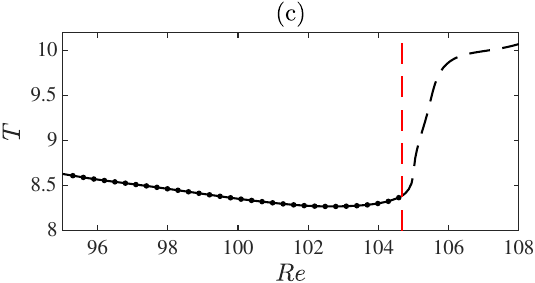}\\
     \vspace{0.5cm}
	\includegraphics[scale=.8]{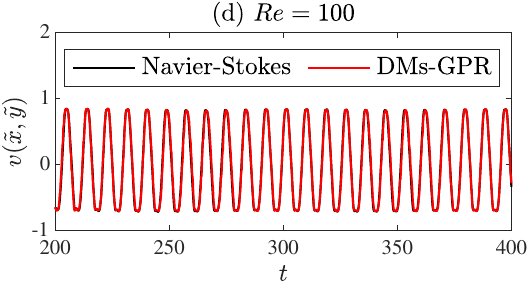}
    \hspace{0.25cm}\includegraphics[scale=.8]{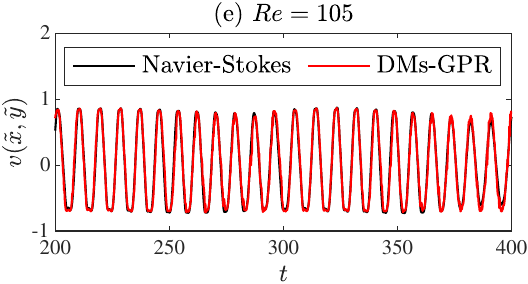}\\
    \vspace{0.5cm}
    \includegraphics[scale=.8]{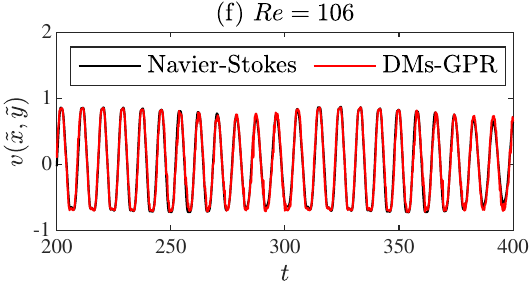}
    \hspace{0.25cm}
    \includegraphics[scale=.8]{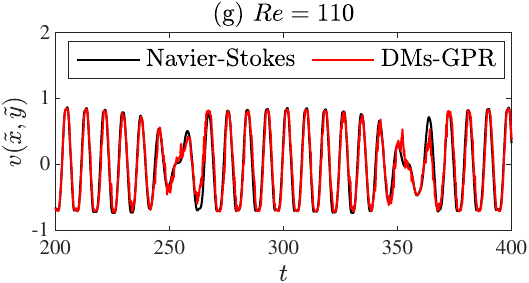}\\
	\caption{Bifurcation diagram of the DMs--GPR reduced-order model of the fluidic pinball, obtained by limit-cycle continuation in MATCONT and visualized in the $(Re,\tilde{a}_1,\tilde{a}_2)$ embedded space (a) and in the $(Re,\tilde{a}_1)$ plane (b). Along the branch of stable limit cycles (solid black curves), a Neimark--Sacker bifurcation is detected at $Re_{ns}=104.67$ (red curves in panels (a)--(b)), from which a branch of unstable limit cycles emerges (dashed curves). The variation of the limit-cycle period $T$ along the branch is reported in panel (c). Panels (d)--(g): comparison between the Navier--Stokes velocity $v$ and the DMs--GPR prediction reconstructed in the high-dimensional space at the location $(\tilde{x},\tilde{y})=(7,0)$ for different Reynolds numbers: $Re=100$ (d), $Re=105$ (e), $Re=106$ (f), and $Re=110$ (g).}
	\label{fig:PIN_DNS_vs_DMs}
\end{figure}

The resulting continuation branch is reported in Fig.~\ref{fig:PIN_DNS_vs_DMs}(a), where the family of limit cycles is visualized in the embedded space $(Re,\tilde{a}_1,\tilde{a}_2)$, and in Fig.~\ref{fig:PIN_DNS_vs_DMs}(b), which shows a section of the bifurcation diagram in the $(Re,\tilde{a}_1)$ plane. Along the branch, stable periodic solutions are represented by solid black curves. A Neimark--Sacker bifurcation is detected at $Re_{ns}=104.67$, highlighted by the red curves in Fig.~\ref{fig:PIN_DNS_vs_DMs}(a)--(b). From this point, a branch of unstable limit cycles emerges (dashed curves), indicating the loss of stability of the primary periodic orbit and the onset of quasi-periodic dynamics.

Along the continuation branch, the Floquet multipliers of the five-dimensional DMs-based ROM are monitored. Their evolution is reported in Fig.~\ref{fig:PIN_poincare}(c). At the Neimark--Sacker bifurcation point, the modulus of a complex-conjugate pair of Floquet multipliers, $f_7$ and $f_8$, associated with transverse perturbations of the primary limit cycle, crosses the unit circle. This crossing determines the loss of stability of the limit cycle and the onset of quasi-periodic motion on an invariant torus.

The values of the Floquet multipliers at the bifurcation point, together with the corresponding period of the limit cycle, are reported in Table~\ref{tab:floquet_ns}. Apart from the complex-conjugate pair $f_7$ and $f_8$ discussed above, the remaining Floquet multipliers are real and satisfy $f_1<1$ and $f_2<1$, indicating that perturbations along the corresponding directions are asymptotically stable. In addition, $f_9=1$ corresponds to the neutral direction associated with the phase invariance of the limit cycle, as consistently observed also in the simpler single-cylinder configuration (see Fig.~\ref{fig_bifurcation_diagram} in Section~\ref{subsec:res_cyl}). The variation of the limit-cycle period along the branch is reported in Fig.~\ref{fig:PIN_DNS_vs_DMs}(c). It exhibits a mild decrease along the stable branch and then increases in the vicinity of the bifurcation, consistently with the deformation of the periodic solution as the secondary instability leading to quasi-periodic dynamics, through low-frequency modulation, is approached.

Finally, by addressing the pre-image problem via the $k$-NN algorithm (see Section~\ref{subsubsec:PDMs}), the DMs--GPR reduced-order solutions are lifted back to the original high-dimensional physical space, allowing a direct comparison with the Navier--Stokes solutions. Such a comparison is shown in Fig.~\ref{fig:PIN_DNS_vs_DMs}(d)--(g), for the same values of $Re$ previously reported in Fig.~\ref{fig:PIN_torus}, both in the pre- and post-bifurcation regimes. The comparison is performed at a representative spatial location within the flow, namely $(\tilde{x},\tilde{y})=(7,0)$. A very good agreement can be observed for all values of $Re$, both in the periodic and quasi-periodic regimes, with slight deviations appearing only at $Re=110$ (Fig.~\ref{fig:PIN_DNS_vs_DMs}(g)), which corresponds to a regime approaching chaotic dynamics.

These results demonstrate that the DMs-based ROM not only captures the qualitative transition from periodic to quasi-periodic dynamics, but also quantitatively reproduces the underlying bifurcation structure of the pinball flow, including the correct stability properties of the limit cycle and the onset of the Neimark--Sacker bifurcation. 

We emphasize that such behavior cannot be reproduced using POD-based ROMs. In particular, POD-ROMs fail to provide a consistent prediction of the secondary instability and the associated torus birth, thus highlighting the importance of the nonlinear manifold representation induced by DMs for accurately describing the flow dynamics beyond the primary vortex-shedding regime.

\section{Conclusions}
\label{sec:conclusions}

We have presented a four-stage, data-driven framework that combines manifold learning and machine learning modelling to perform numerical analysis of the Navier–Stokes equations. A central outcome of this work is that  ROMs built through parsimonious Diffusion Maps \cite{dsilva2018parsimonious,DellaPia_diffusion_2024} enable accurate and computationally efficient numerical analysis of complex fluid flows, including demanding bifurcation and stability computations associated with strongly nonlinear dynamics such as Neimark--Sacker bifurcations. In particular, the use of nonlinear manifold learning allows the constructed ROMs to capture the intrinsic minimal dimensionality of the underlying dynamics, which is essential for reliable continuation and stability analysis. This, in turn, highlights the need to move beyond POD, which remains the standard choice in much of CFD but may fail in such demanding numerial analysis tasks, especially for secondary bifurcations, or require additional modes that do not correspond to the parsimonious normal-form dynamical dimension. As a result, POD-based reductions may lose both dynamical-systems and numerical-analysis insight, whereas parsimonious DMs-based ROMs provide a more faithful and effective foundation for advanced reduced-order modelling and analysis. The methodology was demonstrated on three representative two-dimensional flows of increasing dynamical complexity: the cylinder wake, the planar sudden-expansion channel flow, and the fluidic pinball. In all three cases, the proposed framework enabled accurate stability analysis in addition to bifurcation identification. For the cylinder wake, we performed a complete stability analysis through the computation of Floquet multipliers of the learned limit cycles. For the sudden-expansion flow, we reconstructed the full bifurcation diagram within a single ROM, capturing two coexisting stable attracting solutions, whereas previous studies were limited to continuing one stable branch at a time~\cite{Pichi, HESS2019379, pintore2021efficient}. Importantly, as we have shown, for the fluidic pinball, the parsimonious DMs-based ROM accurately detects the Neimark--Sacker bifurcation, capturing the onset of quasi-periodic dynamics through the birth of an invariant torus, and the associated stability properties, thus allowing the continuation of unstable limit cycles beyond this bifurcation through the arsenal of numerical bifurcation-analysis toolkit, such as MATCONT. By contrast, POD-based ROMs did not reproduce this behavior reliably or even failed, especially in the context of secondary bifurcations.

In this direction, future work will focus on extending the framework to more complex bifurcation scenarios. Of particular interest in Navier--Stokes PDEs are codimension-two and global bifurcations, which generate fundamental qualitative changes in the global phase-space structure and underlie the onset of turbulence. Exploring broader parameter regimes associated with these transitions remains a major challenge, as well as the identification of low-dimensional attracting manifolds and the corresponding normal forms for global bifurcations.

\backsection[Acknowledgements]{A.D.P. acknowledges the CINECA award under the ISCRA initiative, for the availability of high-performance computing resources and support. C. S. acknowledges partial support from the PNRR MUR, projects PE0000013-Future Artificial Intelligence Research-FAIR \& CN0000013 CN HPC - National Centre for HPC, Big Data and Quantum Computing, Gruppo Nazionale Calcolo Scientifico-Istituto Nazionale di Alta Matematica (GNCS-INdAM).}


\backsection[Declaration of interests]{The authors report no conflict of interest.}


%


\appendix
\section{Gaussian Process Regression: Training and Implementation}
\label{app:insights}

\begin{figure}[]
	\centering
	\includegraphics[scale=0.9]{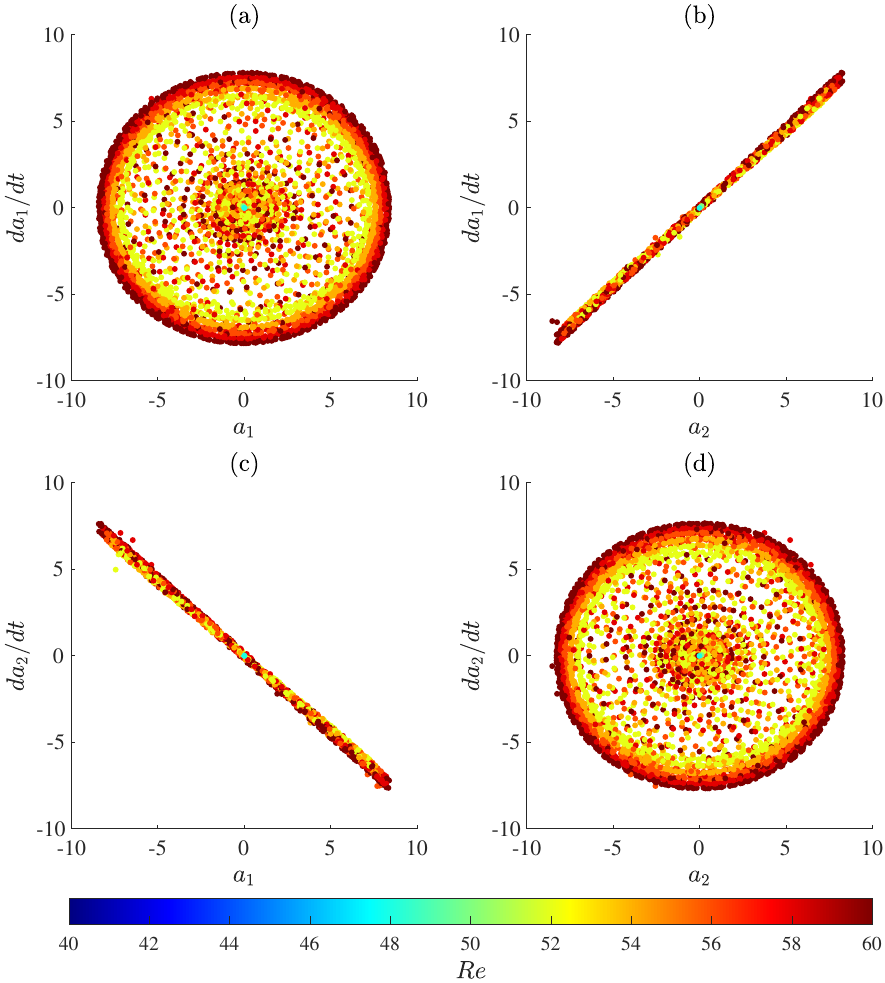}
	\caption{Numerical data set used to train the POD-GPR model of the circular cylinder flow latent dynamics in the $(Re,a_1,a_2)$ space, colored by Reynolds number $Re$.}
	\label{fig:dataset_training_CYL}
\end{figure}

\begin{figure}[]
	\centering
	\includegraphics[scale=0.9]{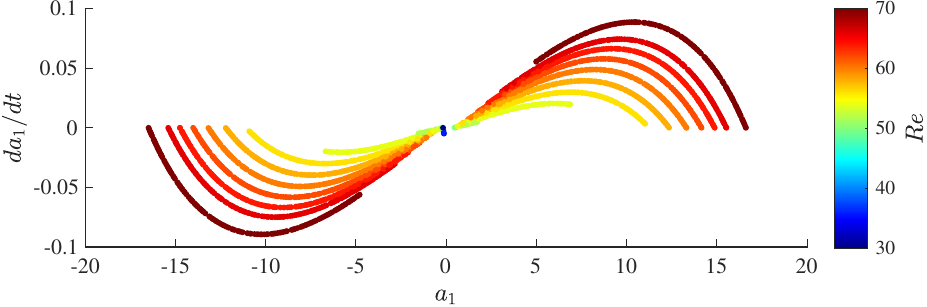}
	\caption{Numerical data set used to train the POD-GPR model of the sudden-expansion channel flow latent dynamics in the $(Re,a_1)$ space, colored by Reynolds number $Re$.}
	\label{fig:dataset_training_BIF}
\end{figure}

\begin{figure}[]
	\centering
	\includegraphics[scale=0.94]{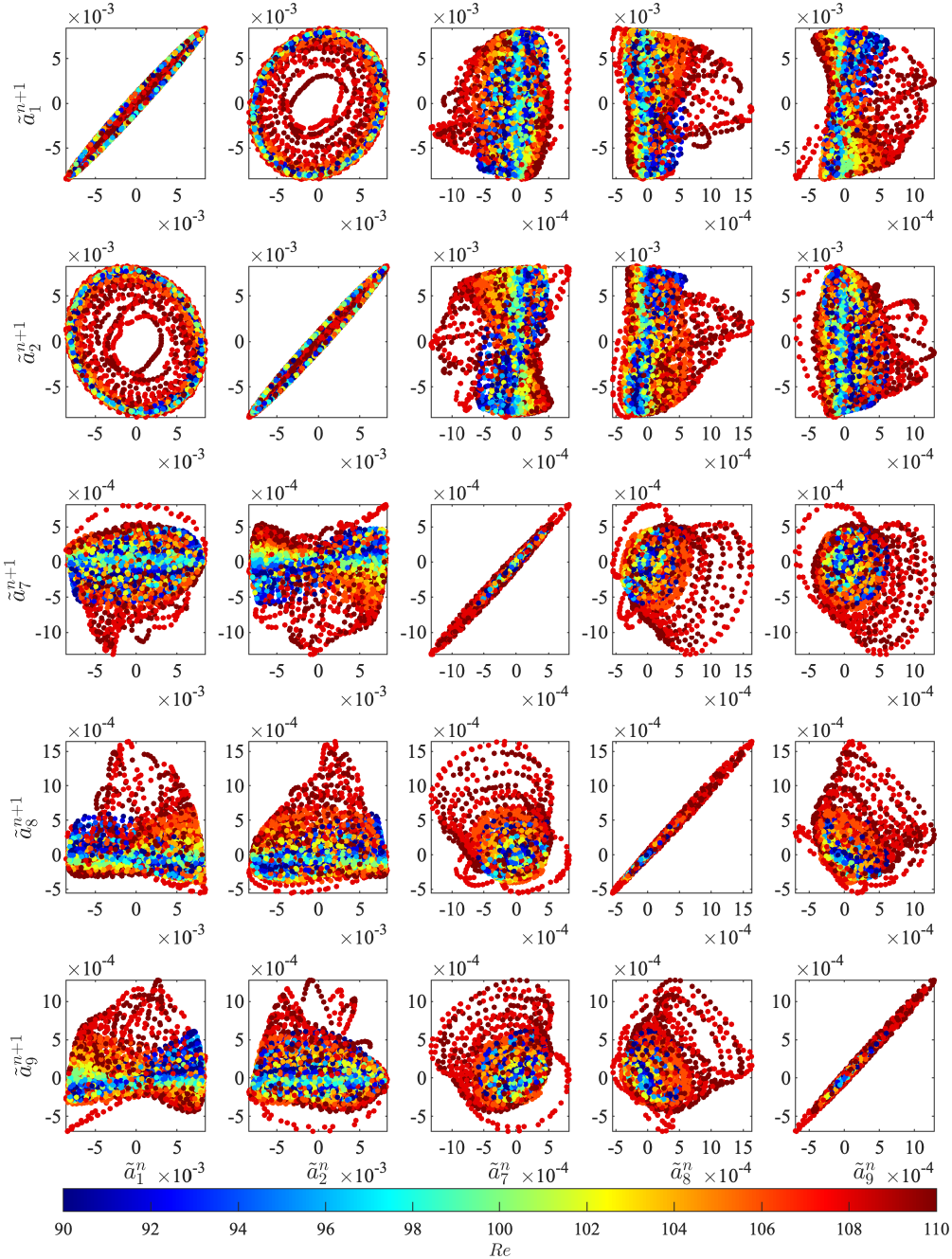}
	\caption{Numerical data set used to train the DMs-GPR model of the pinball flow latent dynamics in the reduced parsimonious DMs coordinates $\tilde{a}_i$ ($i=1,2,7,8,9$), colored by Reynolds number $Re$. Here, $\tilde{a_i}^{n} \equiv \tilde{a_i}(t^n)$ and $\tilde{a_i}^{n+1} \equiv \tilde{a_i}(t^n + T_p)$, being $T_p=0.2$ the time horizon of the prediction.}
	\label{fig:dataset_training_PIN}
\end{figure}

The numerical datasets employed to train the surrogate Gaussian Process regression (GPR) models of the latent dynamics are presented in Figs.~\ref{fig:dataset_training_CYL}--\ref{fig:dataset_training_PIN} for the circular cylinder, sudden-expansion channel, and fluidic pinball configurations, respectively. In all cases, the available data were randomly partitioned into training (60 \%) and testing (40 \%) subsets and subsequently used to identify the reduced dynamical operators within the MATLAB environment according to Eqs.~\eqref{eq:DNN_a1}--\eqref{eq:GPR_tilde_a}.
The time evolution of the reduced coordinates $a_i$ ($i = 1, \dots, d$) was then obtained by integrating the learned latent dynamics. For the cylinder and sudden-expansion channel configurations, a high-order Runge–Kutta scheme was applied to the continuous-time model in Eq.~\eqref{eq:DNN_a1}. In contrast, for the fluidic pinball case, the dynamics were advanced directly by time-marching the discrete evolution map defined in Eq.~\eqref{eq:GPR_tilde_a}.
For the first two benchmarks, the time derivatives $d\mathbf{a}/dt$ in Eq.~\eqref{eq:DNN_a1} were approximated using finite differences with a time step $\Delta t = 0.2$. The same value, $T_p = 0.2$, was adopted to construct the discrete map in Eq.~\eqref{eq:GPR_tilde_a} for the third benchmark.


As discussed in Section~\ref{subsec:basis}, POD is able to correctly identify the intrinsic dimensionality of the data sets for the cylinder ($d=2$) and for the sudden-expansion channel flow ($d=1$). In both cases, the geometry of the training data in the latent space is relatively simple: smooth two- and one-dimensional manifolds parametrized by the Reynolds number and the leading modal amplitudes. This structural simplicity is reflected in Figs.~\ref{fig:dataset_training_CYL} and~\ref{fig:dataset_training_BIF}, where the data organize along low-curvature surfaces, consistent with the presence of a single primary instability (Andronov--Hopf for the cylinder and pitchfork for the channel flow). The smoothness and low curvature of these manifolds facilitate the learning of accurate reduced evolution operators in POD linear subspaces.

In contrast, the training data for the pinball configuration (Fig.~\ref{fig:dataset_training_PIN}) exhibit a significantly richer and more intricate geometric structure. In the $d=5$ parsimonious DMs embedding coordinates $\tilde{a}_i$ ($i=1,2,7,8,9$), the data lie on curved manifolds with visible twisting and folding patterns, reflecting the coexistence and interaction of multiple time scales. These complex patterns are direct manifestations of the secondary instability that leads to quasi-periodic dynamics and the birth of an invariant torus. The latent manifold is no longer well approximated by a linear subspace; rather, it possesses a nontrivial geometry and topology that require nonlinear parametrization. This observation further justifies the use of DMs for this configuration: linear approaches such as POD would not provide a geometrically consistent embedding, nor reliably identify the intrinsic dimensionality necessary to capture the Neimark--Sacker bifurcation and the associated torus dynamics. The DMs coordinates, instead, yield a smooth and dynamically meaningful embedding on which the latent evolution can be effectively learned.

\begin{figure}[] 
\centering \includegraphics[scale=0.8]{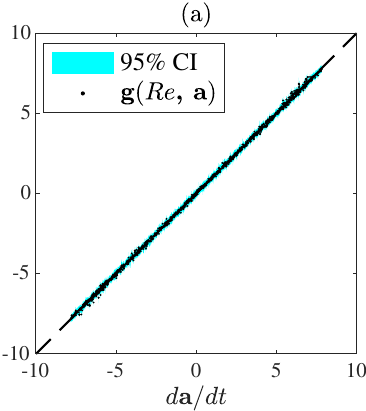} \includegraphics[scale=0.8]{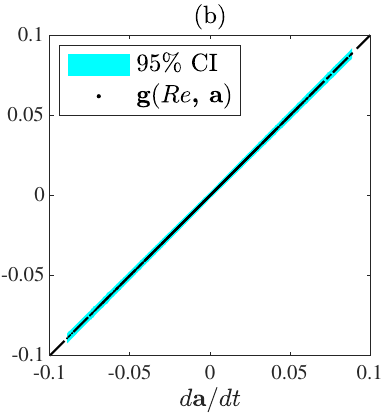} \includegraphics[scale=0.82]{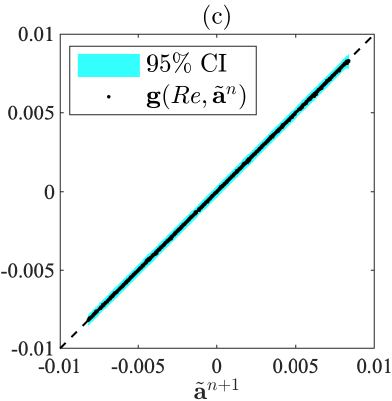} \caption{Uncertainty quantification of the GPR reduced-order models on the testing data set for the cylinder (a), channel (b) and pinball (c) flow configurations. For given input data, the model predictions (i.e., $d\textbf{a}/dt$ for the first and second cases, $\tilde{\textbf{a}}^{n+1}$ for the third case) are shown both in terms of expected (mean) values (black dots) and 95\% CI (confidence intervals, cyan band). The black dashed line in all panels denotes the quadrant bisector.} \label{fig:uncertainty} 
\end{figure}

\begin{figure}[]
	\centering
	\includegraphics[scale=0.85]{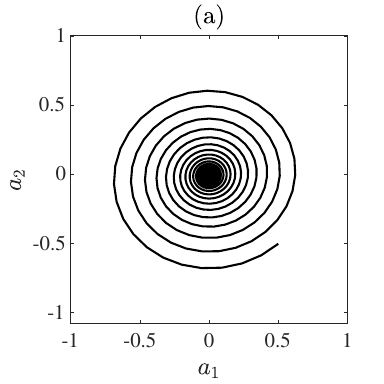}
	\hspace{1.0cm}\includegraphics[scale=0.85]{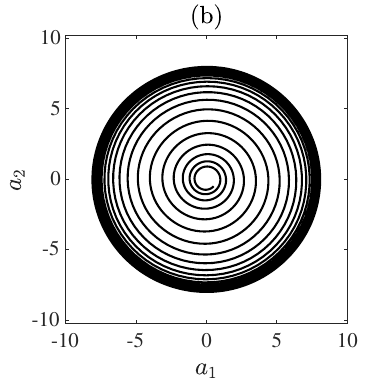}
	\caption{Reduced-order dynamics of the cylinder flow as predicted by time integration of the POD-GPR model by variation of the Reynolds number: $Re=45$ (a); $Re=55$ (b).}
	\label{fig:DNN_dynamics}
\end{figure}

For the numerical integration of the ROMs presented in Sections~\ref{subsec:res_cyl} and~\ref{subsec:res_hannel}, we considered only the expected (mean) value of the time-derivative and the solution operators predicted by the GPR models (Eqs.~\eqref{eq:DNN_a1}--\eqref{eq:GPR_tilde_a}). The predictive accuracy and associated uncertainty levels are illustrated in Fig.~\ref{fig:uncertainty} for the three configurations. For given inputs $(Re,\mathbf{a}) \equiv (Re,a_1,\dots,a_d)$, the GPR model predicts both the expected value of the reduced operator (black dots in Fig.~\ref{fig:uncertainty}) and its standard deviation, from which the 95\% confidence intervals (cyan bands) are computed. For the cylinder and channel configurations, the learned models approximate the continuous-time dynamics $d\mathbf{a}/dt$, whereas for the pinball case the model predicts the discrete-time evolution map $\tilde{\mathbf{a}}^{n+1} \equiv \tilde{\mathbf{a}}(t^n + T_p)$ as a function of the input data $(Re,\tilde{a}_1^{\,n},\dots,\tilde{a}_d^{\,n})$. Note that, here as in Section~\ref{sec:results}, the DMs coordinates are denoted by $\tilde{\mathbf{a}}$ when presented together with the POD coordinates $\mathbf{a}$.

On the testing data sets, the maximum standard deviation was found to be $2.03\%$ (Fig.~\ref{fig:uncertainty}(a)), $3.04\%$ (Fig.~\ref{fig:uncertainty}(b)), and $3.22\%$ (Fig.~\ref{fig:uncertainty}(c)) of the corresponding mean value for the cylinder, channel, and pinball flows, respectively. These relatively small uncertainty levels, even in the more complex pinball case, confirm that the learned surrogate normal-form models provide an accurate and robust approximation of the latent Navier--Stokes dynamics across the explored parameter ranges.

\bibliographystyle{unsrt}
\bibliography{mybibfile}

\end{document}